\documentclass[a4paper,12pt]{article}
\pdfoutput=1 

\usepackage{jheppub} 
\usepackage[T1]{fontenc} 

\usepackage[latin1]{inputenc}
\usepackage[english]{babel}

\usepackage{amssymb,amsthm,cancel,hyperref,graphicx,xcolor}
\usepackage{picinpar,graphicx,xy}
\usepackage[subnum]{cases}
\usepackage{booktabs}
\usepackage{mathrsfs}
\usepackage{amsfonts}
\usepackage{latexsym}
\usepackage{booktabs,graphicx,hyperref,epsfig,multirow}
\usepackage{bm}
\usepackage{comment}
\allowdisplaybreaks
\def\bea{\begin{align}}
\def\eea{\end{align}}
\def\beq{\begin{equation}}
\def\eeq{\end{equation}}
\def\ba{\begin{eqnarray}}
\def\ea{\end{eqnarray}}
\def\be{\begin{equation}}
\def\ee{\end{equation}}
\definecolor{darkgreen}{HTML}{008000}
\newcommand{\sss}{\scriptscriptstyle\rm}
\newcommand{\muf}{\mu_{\rm\sss F}}
\newcommand{\mur}{\mu_{\rm\sss R}}

\newcommand{\Li}{\mathrm{Li}}

\newcommand{\gammae}{\gamma_{\scriptscriptstyle E}}
\newcommand{\Ord}{\mathcal{O}}

\newcommand{\as}{\alpha_s}

\newcommand{\plus}[1]{\left[#1\right]_+}
\def\toinf#1{\mathrel{\mathop{\sim}\limits_{\scriptscriptstyle{#1\rightarrow\infty }}}}
\def\tozero#1{\mathrel{\mathop{\sim}\limits_{\scriptscriptstyle{#1\rightarrow0 }}}}
\def\({\left(}
\def\){\right)}
\def\[{\left[}
\def\]{\right]}
\def    \hepph  #1 {{\tt hep-ph/#1}}
\def    \hepex  #1 {{\tt hep-ex/#1}}
\long\def\symbolfootnote[#1]#2{\begingroup%
\def\thefootnote{\fnsymbol{footnote}}\footnote[#1]{#2}\endgroup}

\numberwithin{equation}{section}
\def\lapprox{\lower .7ex\hbox{$\;\stackrel{\textstyle <}{\sim}\;$}}
\def\gapprox{\lower .7ex\hbox{$\;\stackrel{\textstyle >}{\sim}\;$}}

\renewcommand{\(}{\left(}
\renewcommand{\)}{\right)}

\newcommand{\Ca}{C_{\rm\sss A}}
\newcommand{\Cf}{C_{\rm\sss F}}
\newcommand{\Nf}{n_f}

\newcommand{\Gf}{G_{\sss F}}

\newcommand{\lchi}{\lambda_{\chi}}
\newcommand{\lN}{\lambda_{N}}
\newcommand{\pt}{p_{\rm\sss T}}

\graphicspath{{Images/}}
\begin{document}
\begin{flushleft}
\begin{figure}[h]
\includegraphics[width=.2\textwidth]{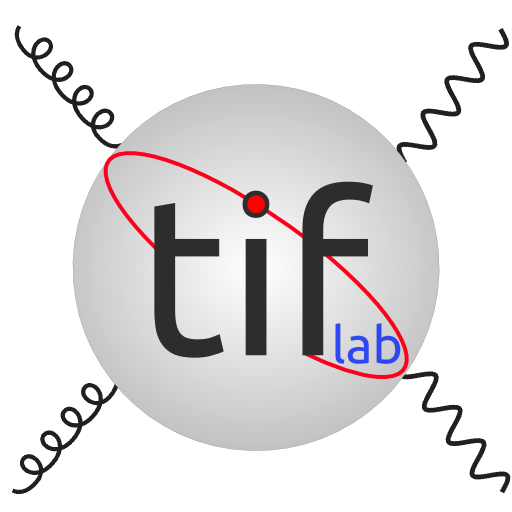}
\end{figure}
\end{flushleft}
\vspace{-5.0cm}
\begin{flushright}
TIF-UNIMI-2016-11
\end{flushright}

\vspace{2.0cm}

\begin{center}
{\Large \bf Combined threshold and transverse momentum resummation for
  inclusive observables}\\
\end{center}

\vspace{1.3cm}

\begin{center}
Claudio Muselli$^1$, Stefano Forte$^1$, Giovanni Ridolfi$^2$. \\
\vspace{.3cm}
{\it
{}$^1$Tif Lab, Dipartimento di Fisica, Universit\`a di Milano and\\
INFN, Sezione di Milano, Via Celoria 16, I-20133 Milano, Italy\\
{}$^2$Dipartimento di Fisica, Universit\`a di Genova and\\
INFN, Sezione di Genova, Via Dodecaneso 33, I-16146 Genova, Italy.\\
}

{\bf \large Abstract}
\end{center}
We present a combined resummation for the  transverse momentum distribution
of a colorless final state in perturbative QCD,
expressed as a function of transverse momentum $\pt$ and the
scaling variable $x$. Its expression satisfies three
requirements: it reduces to standard transverse momentum resummation
to any desired logarithmic order in the limit $\pt\to 0$ for fixed
$x$, up to power suppressed corrections in $\pt$; it reduces to
threshold resummation to any desired logarithmic order in the limit
$x\to 1 $ for fixed $\pt$, up to power suppressed correction in $1-x$;
upon integration over transverse momentum it reproduces the
resummation of the total
cross cross at any given logarithmic order in the threshold
$x\to1$ limit, up to power suppressed correction in $1-x$. Its main ingredient, and our main new result, 
is a modified form of transverse momentum
resummation, which leads to threshold resummation upon integration
over $p_T$, and for which we provide a simple
closed-form analytic expression in Fourier-Mellin $(b,N)$ space. We
give explicit coefficients up to NNLL order for the
specific case of Higgs production in gluon fusion in the effective
field theory limit. Our result allows for a systematic improvement of
the transverse momentum distribution through threshold resummation
which holds for all $\pt$, and elucidates the relation between
transverse momentum resummation and threshold resummation at the
inclusive level, specifically by providing within perturbative QCD 
a simple derivation of the
main consequence of the
so-called collinear anomaly of SCET.

\clearpage
\tableofcontents
\clearpage

\section{Introduction}\label{sec:introduction}

Transverse momentum $\pt$ distributions of inclusive observables, such as
the Higgs or gauge-boson production cross sections, receive
logarithmically enhanced corrections in perturbative QCD,
related to infrared and collinear radiation. While infrared
radiation is also collinear, the opposite is not true. As a
consequence, the all-order resummation of transverse momentum
logarithms (transverse momentum resummation, henceforth) contains more
information than  the resummation of infrared threshold logarithms 
(threshold resummation, henceforth) of the total cross section,
integrated over transverse momentum (total cross section, henceforth):
it should therefore be possible to 
recover the latter from the former. On the other hand, the transverse
momentum distribution in the threshold limit for finite $\pt$  necessarily has a
different logarithmic structure than the total cross section because
there must be at least one non-soft emission recoiling against the
final state. Therefore, the soft and collinear limits do not commute:
if the collinear limit is taken first, extra logs are present which
are not there when the soft limit is taken at finite $\pt$, and conversely.

All-order techniques have been available since a long time for
transverse momentum resummation~\cite{Collins:1984kg},
threshold resummation of 
total cross sections~\cite{Catani:1989ne,Sterman:1986aj}, and more recently also for  threshold resummation
of transverse 
momentum distributions~\cite{deFlorian:2005fzc}. Their combination,
however, is nontrivial because of the non-commutative nature of the
soft and collinear limits. It is the purpose of this paper to provide
such a combination. In particular, our goal is to construct an
all-order resummed expression for transverse momentum distributions
which has the following properties:
\begin{enumerate}
\item at small $\pt$ it reproduces  transverse momentum
  resummation to some fixed logarithmic accuracy;
\item in the threshold $x\to 1$ limit for finite $\pt$ it reproduces
  threshold resummation also to some given accuracy;
\item upon integration over $\pt$ it leads to a total cross section
  which reproduces threshold resummation to some logarithmic
  accuracy.
\end{enumerate}

Some previous attempts to combined resummation have been presented,
but they fail to simultaneously satisfy all of these
criteria. Specifically, in 
Ref.~\cite{Laenen:2000ij} a joint threshold and transverse momentum
resummation has been presented which satisfies criteria~1 and~3, but
fails to satisfy criterion~2. This resummation, originally presented
up to next-to-leading logarithmic (NLL) accuracy, was very recently
extended to NNLL in the specific case of Drell-Yan
production~\cite{Marzani:2016smx}. Also, the non-commutativity of the
limits was recently recognized in
Refs.~\cite{Li:2016axz,Lustermans:2016nvk}, and the two resummations were combined at the
differential level using SCET through a suitable  tuning of
  the scales involved in such a way that the relevant limits do not
  mix. This leads to a result which satisfies criteria~1 and~2, but 
  fails to satisfy criterion~3: the integral over
  $\pt$ does not
  reproduce the threshold resummation for inclusive cross section. 

A crucial step towards the construction of our result is a 
careful analysis of the way
phase space factorizes in the two limits we are interested in. Indeed,
it turns out that in order to obtain either of the two resummations
that we wish to combine, a different factorization of phase space is
necessary, and also, that the threshold and small $\pt$ limits do not
commute because of the structure of phase space. 
This
phase-space analysis will be presented in Sect.~\ref{sec:phasespace},
  after an introductory Sect.~\ref{sec:notations}, 
in which various known individual
  resummed results are summarized and notation is introduced.
Our combined resummed results will then be presented in
Sect.~\ref{sec:joint}, with specific reference to the case of Higgs  
production, with conclusions drawn in 
Sect.~\ref{sec:conclusion}.

\section{Factorization and resummation of transverse momentum distributions}
\label{sec:notations}   

\subsection{Notations and kinematics}\label{subsec:kin}
We discuss the transverse momentum  distribution of a colorless
object which we shall refer to as ``Higgs'' for the sake of brevity,
though it could
just as well be e.g.\ a gauge boson. 
The factorized perturbative QCD expression for the Higgs $\pt$ 
distribution at a hadron collider takes the form
\begin{align}
\label{eq:conv1}
\frac{d\sigma}{d\xi_p}\(\tau,\xi_p,M^2\)&=
\sum_{i,j}\int_{\frac{\tau}{\(\sqrt{1+\xi_p}-\sqrt{\xi_p}\)^2}}^1 dx_1\,
\int_{\frac{\tau}{x_1\(\sqrt{1+\xi_p}-\sqrt{\xi_p}\)^2}}^1 dx_2\,
f_i\(x_1,\muf^2\)\,f_j\(x_2,\muf^2\)\notag\\
&\frac{d\bar{\sigma}_{ij}}{d\xi_p}\(\hat \tau,\xi_p,\as\(\mur^2\),\muf^2\),
\end{align} 
where $\frac{d\sigma}{d\xi_p}$ and $\frac{d\bar\sigma_{ij}}{d\xi_p}$ are
the hadronic and partonic distributions, written in
terms of the dimensionless transverse momentum variable
\begin{align}\label{xidef}
\xi_p = \frac{\pt^2}{M^2}
\end{align}
and the scaling variables
\begin{align}\label{eq:hadtau}
\tau &= \frac{M^2}{s};\\ \label{eq:parttau}
\hat\tau &= \frac{M^2}{\hat s},
\end{align}
where $M^2$ is the invariant Higgs mass, and $s$ and
\begin{align}\label{partcm}
\hat s= x_1x_2 s 
\end{align}
are respectively the hadronic and partonic center-of-mass squared
energies. The
partonic differential cross section is perturbatively determined in terms of the
strong coupling $\alpha_s(\mu^2_R)$  at a renormalization scale
$\mu_R$  and a factorization scale $\mu_F$,
and $f_i\(x_1,\muf^2\)$ are parton distributions.

Equation~\eqref{eq:conv1} can be rewritten as a convolution by defining
\begin{align}\label{tauprimedef}
\tau'=\frac{Q^2}{s},
\end{align}
where 
\beq\label{scaldef}
Q^2=\(\sqrt{M^2+\pt^2}+\pt\)^2.
\eeq
Indeed, we then get
\begin{align}
\label{eq:conv2}
\frac{d\sigma}{d\xi_p}\(\tau,\xi_p,M^2\)=\tau'\sum_{ij}\int_{\tau'}^1 \frac{dx}{x}\, \mathcal{L}_{ij}\(\frac{\tau'}{x},\muf^2\)\frac{1}{x}\frac{d\hat{\sigma}_{ij}}{d\xi_p}\(x,\xi_p,\as\(\mur^2\),\muf^2\),
\end{align}
where 
the parton luminosity is defined in the usual way as
\beq
\mathcal{L}_{ij}\(x,\muf^2\)=\int_{x}^1 \frac{dy}{y} f_i\(y,\muf^2\)f_j\(\frac{x}{y},\muf^2\),
\eeq
and 
\beq
\label{eq:Cpthadr}
\frac{d\hat{\sigma}_{ij}}{d\xi_p}\(x,\xi_p,\as(\mur^2),\muf^2\)=\frac{d\bar{\sigma}_{ij}}{d\xi_p}\(\frac{x}{\(\sqrt{1+\xi_p}+\sqrt{\xi_p}\)^2},\xi_p,
\as(\mur^2),\muf^2\),
\eeq
so that the partonic scaling variable in Eq.~(\ref{eq:conv2}) is 
\begin{equation}\label{eq:partx}
x=\frac{Q^2}{\hat s}=\frac{M^2}{\hat s}\(\sqrt{1+\xi_p}+\sqrt{\xi_p}\)^2
\end{equation}

The scale variable $Q^2$ Eq.~(\ref{scaldef}) has a
straightforward physical interpretation, as the minimum allowed value of the  
invariant mass of the final state at fixed transverse momentum.
Consequently, the threshold limit
corresponds to the limit $\tau^\prime\to1$ of the scaling 
variable Eq.~(\ref{tauprimedef}), and at the partonic level it
corresponds to the integration region $x\sim1$ in
Eq.~(\ref{eq:conv2}).

\subsection{Factorization}
\label{subsec:fact}

Threshold resummation is most naturally performed on the Mellin
transform of the cross section.
Performing a  Mellin transform with respect to $\tau'$ the convolution
Eq.~\eqref{eq:conv2} factorizes. Indeed, defining
\begin{align}
\label{eq:defMellinhadro}
\frac{d\sigma}{d\xi_p}\(N,\xi_p\)&=\int_0^1 d\tau'\,\tau'\,^{N-1}
\frac{d\sigma}{d\xi_p}\(\tau,\xi_p\);\\
\label{eq:defMellin}
\frac{d\hat{\sigma}_{ij}}{d\xi_p}\(N,\xi_p\)&=\int_0^1 dx\, x^{N-1} \frac{d\hat{\sigma}_{ij}}{d\xi_p}\(x,\xi_p\),
\end{align}
where with a slight abuse of notation we are using the same symbol for
the cross section and its Mellin transform,
we get
\beq
\label{eq:conv3}
\frac{d\sigma}{d\xi_p}\(N,\xi_p,M^2\)=\sum_{ij} \mathcal{L}_{ij}\(N+1,\muf^2\) \frac{d\hat{\sigma}_{ij}}{d\xi_p}\(N,\xi_p,\as\(\mur^2\),\muf^2\).
\eeq

It is important to observe that, because the scale $Q^2$
Eq.~(\ref{scaldef}) and consequently the scaling variable
$\tau^\prime$ Eq.~(\ref{tauprimedef}) both depend on $\pt$,
the Mellin transforms of the total hadronic and partonic cross sections 
\begin{align}
\label{eq:totdefh}
\sigma(\tau) 
&= \int_0^{\frac{(1-\tau)^2}{4\tau}} d\xi\,
\frac{d\sigma}{d\xi}(\tau,\xi)
\\
\label{eq:totdefp}
\hat\sigma_{ij}(\hat \tau) 
&= \int_0^{\frac{(1-\hat \tau)^2}{4\hat \tau}} d\xi\,\frac{d\bar\sigma_{ij}}{d\xi}(\hat\tau,\xi)
\end{align}
are not just the
integrals over $\pt$ of the Mellin transforms
Eq.~(\ref{eq:defMellinhadro},\ref{eq:defMellin}) 
of the respective transverse momentum
distributions. Rather, defining the Mellin transforms of the total
hadronic and partonic cross sections
\begin{align}\label{eq:tothadmel}
\sigma\(N\)&=\int_0^1 d\tau\,\tau^{N-1} \sigma\(\tau\)\\ 
\label{eq:totpartmel}
\hat{\sigma}_{ij}\(N\)&=\int_0^1 d\hat\tau\,\hat\tau^{N-1} \hat{\sigma}_{ij}\(\hat\tau\)
\end{align}
which factorize the  cross section as
\beq\label{eq:totfact}
\sigma\(N,\as\(\mur^2\),\muf^2\)=
\sum_{ij}\mathcal{L}_{ij}\(N+1,\muf^2\)
\hat{\sigma}_{ij}\(N,M^2\),
\eeq
and using Eqs.~(\ref{eq:partx},\ref{eq:totdefp}) we get
\beq\label{eq:melrel}
\hat{\sigma}_{ij}\(N\)=\int_0^{\infty}d\xi_p\, \(\sqrt{1+\xi_p}-\sqrt{\xi_p}\)^{2N} \frac{d\hat{\sigma}_{ij}}{d\xi_p}\(N,\xi_p\).
\eeq
It is very important to observe that here and everywhere henceforth we
will assume that the  Mellin transform is defined for transverse
momentum distributions according to
Eqs.~(\ref{eq:defMellinhadro}-\ref{eq:defMellin}) (with the partonic
differential cross section
$\frac{d\hat{\sigma}_{ij}}{d\xi_p}\(N,\xi_p\)$ given by Eq.~(\ref{eq:Cpthadr})), and for total
cross sections according to
Eqs.~(\ref{eq:tothadmel}-\ref{eq:totpartmel}). They are thus related
by Eq.~(\ref{eq:melrel}), as a consequence of the fact that the
integral transformation kernel differs between the two cases.

Transverse momentum resummation is best performed on the
Fourier transform of the transverse momentum distribution. 
This is defined as
\beq
\label{eq:usualFouriertransform}
\frac{d\hat{\sigma}_{ij}}{d\xi_p}\(N,b\)=\frac{1}{M^2}\int d^2 p_{\sss T}\, e^{i \vec{b} \cdot \vec{p}_{\sss T}} \frac{d\hat{\sigma}_{ij}}{d\xi_p}\(N,\xi_p\),
\eeq
which, performing the angular integration, becomes
\beq
\label{eq:usualFouriertransform2}
\frac{d\hat{\sigma}_{ij}}{d\xi_p}\(N,b\)=\pi \int_0^\infty d\xi_p\,  
J_0\(b p_{\sss T}\)\frac{d\hat{\sigma}_{ij}}{d\xi_p}\(N,\xi_p\),
\eeq
where $J_0$ is the Bessel function.

It is interesting to observe that we are free to take a simultaneous Mellin
and  Fourier transform thanks to the fact that the Mellin
transform is performed, according to Eq.~(\ref{eq:defMellinhadro}),
with respect to the scaling variable $\tau^\prime$ 
Eq.~(\ref{tauprimedef}). Indeed, the variable
$\tau^\prime$ ranges from
$0 \le \tau^\prime \le1$ for all $\pt$, and $\pt$ ranges from $0\le
\pt\le\infty$ for all $\tau^\prime$. This is to be contrasted to the
the situation in which the kinematics is parametrized by $\pt$ and the
scaling variable $\tau$ Eq.~(\ref{eq:hadtau}). In that case, for fixed
$\pt$, $\tau$  has a $\pt$-dependent upper boundary:
\begin{align}\label{eq:taurange}
0\le\tau\le \left(\sqrt{1+\xi_p}-\sqrt{\xi_p}\right)^2.
\end{align}
Conversely, for fixed $\tau$, $\pt$ has a $\tau$-dependent range, most
easily expressed in terms of the dimensionless variable $\xi_p$:
\begin{align}\label{eq:ptrange}
0\leq\xi_p \leq \frac{\(1-\tau\)^2}{4 \tau}.
\end{align}
Hence, it is not possible to take a Mellin transform with respect to
$\tau$ of the $\pt$ distribution, or a Fourier transform with respect
to $\pt$ at fixed $\tau$, without extending the integration range
outside the physical region. Note, however, that when $\pt\to0$ the
upper limit of $\tau$ Eq.~(\ref{eq:taurange}) tends to one. Equivalently,
\begin{align}\label{eq:xipexp}
\tau^\prime=\tau \left(1+ O(\sqrt\xi_p)\right).
\end{align}
It follows
that 
transverse momentum resummation can be performed using the scaling
variable $\tau$ Eq.~(\ref{eq:hadtau}) up to corrections which are
power-suppressed in the small-$\pt$ limit. However, in order to obtain
results which hold also when $\pt$ is kept finite, as it is our goal,
the scaling variable $\tau^\prime$ 
Eq.~(\ref{tauprimedef}) must be used. 

\subsection{Threshold resummation at large $\pt$}
\label{subsec:threshold}

Threshold resummation for the $\pt$ distribution of colorless final
states was presented in Ref.~\cite{deFlorian:2005fzc}. It is based on
the observation that it is merely a particular case of soft
resummation of a process which at leading-order has more than one
parton in the final state~\cite{Bonciani:2003nt}, such as prompt-photon
production~\cite{Catani:1998tm}, which
corresponds to the particular case in which the mass $M$ which enters
the definition of the scale variable Eq.~(\ref{scaldef}) vanishes.

As discussed  in Ref.~\cite{Catani:1998tm,Bolzoni:2005xn}, soft resummation
in this case is characterized by the fact that there are two different classes
of soft emissions, characterized by two different scales:
 one related to the emission of soft partons (akin to that for
soft resummation in gauge or Higgs boson production) and one related
to emission of partons which are collinear to the parton whose transverse
momentum balances that of the colorless final state. Soft partons, in
turn, can be emitted either collinear to the incoming leg, or at large
angle: the contribution from the latter then depends on $\pt$, as we
shall see more explicitly in Sect.~\ref{sec:phasespacelogN} below.

Threshold resummation of transverse momentum distributions takes the form
\begin{equation}
\label{eq:largeNjoint}
\frac{d\hat{\sigma}_{ij}^{\rm th}}{d\xi_p}\(N,\xi_p,\as\(Q^2\),Q^2\)=\sigma_0\,C_0\(N,\xi_p\)\,g_0\,_{ij}\(\xi_p\)\exp\left[G\(N\)\right] \exp\left[S\(N,\pt\)\right]
\end{equation}
where $Q^2$ is the scale Eq.~(\ref{scaldef}),  and the resummation of soft large-angle emissions is
included in the $\pt$-dependent soft function $S\(N,\pt\)$, while all
other soft and collinear emissions are resummed into the
$\pt$-independent Sudakov exponent  $G\(N\)$. All 
contributions which do not vanish as $N\to\infty$ but are also not
logarithmically enhanced are contained in the matching function
$\sigma_0\,C_0\(N,\xi_p\) g_0\,_{ij}(\xi_p)$ which is a
power series in $\as$, with
$\sigma_0$ the leading-order (LO) inclusive cross section and
$\sigma_0\,C_0\(N,\xi_p\)$ the unresummed 
LO transverse momentum
distribution.

The Sudakov exponent  has the structure
\beq
\label{eq:GN}
G\(N\)=\Delta_i\(N\)+\Delta_j\(N\)+J_k\(N\)
\eeq
where  $i,j$ are the initial partons,  $k$ is the final hard
recoiling parton, and $i,j,k=g,q$. Also,
\begin{align}
\label{eq:deltaJ}
\Delta_i\(N\)&=\int_0^1 dz\,\frac{z^{N-1}-1}{1-z}\int_{Q^2}^{Q^2\(1-z\)^2} \frac{dq^2}{q^2} A_i^{\rm th}\(\as\(q^2\)\)\\
J_k\(N\)&=\int_0^1 dz \, \frac{z^{N-1}-1}{1-z}\int_{Q^2\(1-z\)^2}^{Q^2\(1-z\)} \frac{dq^2}{q^2} A_k^{\rm th}\(\as\(q^2\)\)+B_k^{\rm th}\(\as\(Q^2\(1-z\)\)\)
\end{align}
where $A_i^{\rm th}\(\as\)$ and $B_i^{\rm th}\(\as\)$ are power series
in $\alpha_s$ with numerical coefficients. 

The soft exponent is entirely determined by
the soft anomalous dimension, and thus depends only on the partons
which enter the LO process: at any order in $\as$ it takes the form~\cite{Aybat:2006wq,Gardi:2009qi}
\beq
\label{eq:SN}
S\(N,\xi_p\)=-\int_0^1 dz\, \frac{z^{N-1}-1}{1-z} A^{\rm th}_k\(\as\(Q^2\(1-z\)^2\)\) \ln\frac{\(\sqrt{1+\xi_p}+\sqrt{\xi_p}\)^2}{\xi_p},
\eeq
where $A_k^{\rm th}$ is the so-called cusp anomalous dimension,
i.e. the coefficient of the most singular contribution to the
anomalous dimension as $x\to1$  and $k=g,q$ the
type of the recoiling hard final parton.\footnote{When comparing  the
  form given in Eqs.~(\ref{eq:largeNjoint}-\ref{eq:SN}) of the
  resummation to that given in Eqs.~(12-17) of
  Ref.~\cite{deFlorian:2005fzc} it should be kept in mind that in that
  reference the dependence on $\pt$ is given through a variable $r$
  defined as $r=\sqrt{\frac{\xi_p}{1+\xi_p}}$. Also, the scale
  variable $Q^2$ in that reference is chosen as
  $Q^2=\pt^2\frac{1+r}{r}$, which differs from our choice
  Eq.~(\ref{scaldef}). This difference can be reabsorbed by suitably
  redefining order by order the coefficients of the expansion of  the
  functions $A$ and $B$ and the expression of $S$.}
It follows that $G\(N\)$ and $S\(N,\xi_p\)$ Eqs.~(\ref{eq:GN}-\ref{eq:SN}) can be entirely derived
from knowledge of threshold resummation at the inclusive level;
coefficients 
up to
NNLL are given in Ref.~\cite{Moch:2005ba}. The matching function can be
determined up to N$^k$LL by matching to the fixed N$^k$LO result.

It is important to observe that the leading-order cross section
$\frac{d\hat{\sigma}_{ij}^{\rm LO}}{d\xi_p}\(N,\xi_p\)$ vanishes as
$\frac{1}{\sqrt{N}}$ as $N\to\infty$; correspondingly, the $\tau$-space
  partonic cross section is an ordinary function, rather than a
  distribution, unlike the partonic total  Drell-Yan
  production cross section, and like the heavy quark production
  cross section, as discussed in Ref.~\cite{Muselli:2015kba}.

\subsection{Transverse momentum resummation}
\label{subsec:transverse}
Transverse momentum resummation was first formalized in
Ref.~\cite{Collins:1984kg}. Here we use the form of it given in
Refs.~\cite{Catani:2000vq,Bozzi:2005wk}: the partonic resummed cross section
has the form
\begin{equation}
\label{eq:smallptjoint3}
\frac{d\hat{\sigma}_{ij}^{\rm tr}}{d\xi_p}\(N,\xi_p,\as\(M^2\),M^2\)
=\sigma_0 
\int_0^{\infty} db\,\frac{b}{2}\,J_0\(b \pt\)  H_{ij}\(N,\as\(M^2\)\) S(M,b)+\Ord\(\frac{1}{b}\)
\end{equation}
where  we have chosen $\mu_R^2=\mu_F^2=M^2$, $\sigma_0$ is the LO
inclusive cross section, the hard cross section
$H_{ij}\(N,\as\(M^2\)\)$ is evaluated
at fixed order in perturbation theory, and
the Sudakov form factor $ S(M,b)$ is given by
\begin{equation}\label{eq:sudpt}
 S(M,b)= \exp\left[-\int_{\frac{b_0^2}{b^2}}^{M^2} \frac{dq^2}{q^2}\left[
    A^{\pt}\(\as\(q^2\)\)\ln\frac{M^2}{q^2}+{ B^{\pt}}
\(\as\(q^2\),N\)\right]\right]+\Ord\(\frac{1}{b}\).
\end{equation}
Both $A^{\pt}$ and $B^{\pt}$ and the hard function are series in $\alpha_s$;
$A^{\pt}$ has purely numerical coefficients while all the
$N$-dependence at the exponent is contained in  $B^{\pt}$ which starts at NLL level. Moreover, we factorize the LO inclusive cross section in Eq.~\eqref{eq:smallptjoint3} so that $H_{ij}=1+\Ord\(\as\)$.
The leading logarithmic contribution is
fully determined by leading-order $A^{\pt}$, and each contribution to 
$A^{\pt}$ is enhanced by one logarithmic order in comparison to the
contribution to   $B^{\pt}$  of the same order in $\alpha_s$ because
of the explicit logarithm multiplying it in Eq.~(\ref{eq:sudpt}). An
$N$-independent contribution to  $B^{\pt}$ can be traded for a
contribution to  $A^{\pt}$ with one less power of $\alpha_s$, however,
because $A^{\pt}$ starts to contribute at LL and   $B^{\pt}$ at NLL,
the separation into  $A^{\pt}$ and  $B^{\pt}$  is
uniquely determined, e.g. by 
expanding out Eq.~(\ref{eq:smallptjoint3}) and matching to the
fixed-order computation. The physical meaning of $A^{\pt}$ and
$B^{\pt}$  will become clear once we work out the relation between
transverse momentum resummation and soft resummation at the inclusive
level, and will be discussed in the end of Sect.~\ref{sec:joint}. As a
byproduct, we will derive the relation  between $A^{\pt}$  and
the cusp anomalous dimension  $A^{\rm th}$. The fact that these two
quantities differ was originally found in Ref.~\cite{Becher:2010tm},
where it was shown to follow in a SCET approach 
from a peculiar property of the effective Lagrangian dubbed
``collinear anomaly'': our result will provide a complementary
derivation using standard perturbative QCD techniques.

Note that in Ref.~\cite{Collins:1984kg} the
scale $\mu_F$ at which the parton distributions are evaluated is
chosen  to be a soft scale $\mu_F=\frac{1}{b}$. The difference can be
reabsorbed into the function $B^{\pt}$; the choice made here
leads to a simpler relation between transverse momentum resummation
and soft resummation of the inclusive cross section. 
Because it contains the evolution of the PDFs from the scale
$\frac{1}{b}$ to the scale $M^2$, $B^{\pt}$ is a matrix in  flavor
space, determined in terms of the anomalous dimension matrix. The
 exponential in Eq.~(\ref{eq:smallptjoint3}) should thus be
 interpreted as  path-ordered. Note finally that
there is a certain latitude in defining the function
$H_{ij}\(N,\as\(M^2\)\)$, which includes terms which are not
logarithmically enhanced as $\pt \to0$ and is computed to finite order
in perturbation theory by matching to the finite order result. This
freedom modifies  the Sudakov form factor from NNLL onward, because a
constant (i.e. non-logarithmic) term is of the same logarithmic order
as a NNLL contribution. It can thus be viewed as a resummation scheme
ambiguity.  In the sequel we will adopt the hard resummation scheme
defined in Ref.~\cite{Catani:2013tia}.

\section{Phase space factorization}
\label{sec:phasespace}
It was shown in Ref.~\cite{Contopanagos:1996nh} (and more recently
using effective field theory arguments, see
Ref.~\cite{Becher:2014oda}) that
transverse momentum and threshold resummation can both be derived by
using the renormalization group from a suitable factorization of the
cross section. In Ref.~\cite{Forte:2002ni} this factorization was
proven for inclusive colorless observables in the threshold limit; this
argument was generalized in Ref.~\cite{Bolzoni:2005xn} to the case of
prompt-photon production, whose resummation, as mentioned in
Sect.~\ref{subsec:threshold} above, can be viewed as a particular
case of threshold  resummation at fixed $\pt$. 

The factorization argument of
Refs.~\cite{Forte:2002ni,Bolzoni:2005xn} relies on a kinematical
analysis of phase-space for the process
\beq\label{eq:process}
g\(p_1\)+g\(p_2\) \to \mathcal{S}\(p\)+g\(k_1\)+\dots+g\(k_n\),
\eeq
where $\mathcal{S}$ is a colorless system and $g$ are massless gluons.
 We will now review this argument and show how it relates to the
resummed result Eq.~(\ref{eq:largeNjoint}). This analysis will
elucidate why threshold resummation at fixed $\pt$  only resums a
subset of the soft radiation which is included in transverse momentum
resummation. We will then discuss the different factorization of phase space
which is needed in order to perform transverse momentum
resummation. It will then be clear how to modify this latter factorization by
the inclusion of suitable subleading terms such that the threshold
limit is not spoiled, and in particular threshold
resummation at the inclusive level follows from it, though still not
the full threshold resummation at fixed $\pt$. The simultaneous
transverse momentum and threshold resummation for all $\pt$ will
involve matching these two different forms of factorization, and will
be presented in the next section.

\subsection{Large $N$}
\label{sec:phasespacelogN}

The phase-space factorization of
Refs.~\cite{Forte:2002ni,Bolzoni:2005xn} is based on the iterative
reduction of the $n+1$-body phase  space $d\Phi$ for the process
Eq.~(\ref{eq:process}) into an $n$-body and a two-body
phase space, which eventually leads to expressing it in terms of $n$
two-body phase spaces, connected by integrations over intermediate
virtual particle masses. Physically, this corresponds to iteratively
writing the phase space  in terms of the momentum of the last radiated
particle, and the system containing the previous $n-1$ final-state
ones:
\begin{align}
\label{eq:Phin}
d\Phi\(p_1,p_2;k_1,\dots,k_n,p\)&=\frac{dP_n^2}{2\pi}d\Phi\(p_1,p_2;k_n,P_n\)
\frac{dP_{n-1}^2}{2\pi}d\Phi\(P_n;k_{n-1},P_{n-1}\)\nonumber\\
&\qquad \dots
\frac{dP_2^2}{2\pi}d\Phi\(P_3;k_2,P_2\)d\Phi\(P_2;k_1,p\),
\end{align}
where each of the intermediate particle's invariant mass $P_i^2$ ranges between 
\begin{align}
\label{eq:prange}
M^2\le P^2_i \le  P^2_{i+1}
\end{align}
and $P_{n+1}^2\equiv \hat s$ is the total center-of-mass energy squared.

The result is then simplified taking advantage of the Lorentz 
invariance of each two-body
phase space, in order to rewrite  it in the rest frame of its
incoming momentum $P_i$: in $d=4-2\epsilon$ dimensions 
\begin{align}
d\Phi\(P_i;k_j,P_j\)&=\frac{\(2\pi\)^{2-d}}{4}
\frac{d^{d-1} k_j}{|\vec k_j|\sqrt{|\vec k_j|^2+P_j^2}}\,
\delta\(\sqrt{P_i^2}-\sqrt{|\vec k_j|^2+P_j^2}-|\vec k_j|\)\notag
\\
\label{eq:phi21}
&=\frac{ \(4\pi\)^{2\epsilon-2}}{2}
\(P_i^2\)^{-\epsilon}\(1-\frac{P_j^2}{P_i^2}\)^{1-2\epsilon}
 \(\sin\theta_j\)^{1-2\epsilon} d\theta_j\,d\Omega^j_{2-2\epsilon}
\\
\label{eq:phi22}
&=\frac{\(2\pi\)^{2\epsilon-2}}{8}\(k_{\sss T_j}^2\)^{-\epsilon}
\frac{dk_{\sss T_j}^2\, d\Omega^j_{2-2\epsilon}}
{\sqrt{P_i^2}\sqrt{\frac{P_i^2}{4}\(1-\frac{P_j^2}{P_i^2}\)^2-k_{\sss T_j}^2}},
\end{align}
where the angular integral is written in terms of a 
$(1-2\epsilon)$-dimensional azimuthal integration over
$d\Omega^i_{2-2\epsilon}$, and, equivalently, either a polar
integral over
$\theta_i$ in  Eq.~(\ref{eq:phi21}), or the modulus square of the transverse
momentum   in Eq.~(\ref{eq:phi22}). In the latter case, the
square-root factor in the denominator of  Eq.~(\ref{eq:phi22}) is the
Jacobian related to this new choice of integration variable.

Using this result  the $d$-dimensional phase space Eq.~(\ref{eq:Phin}) becomes
\begin{align}
\label{eq:Phin1}
d\Phi&= \(4\pi\)^{2\epsilon-3}\(P_{n+1}^2\)^{-\epsilon}\(1-\frac{P_n^2}{P_{n+1}^2}\)^{1-2\epsilon} \(\sin\theta_n\)^{1-2\epsilon} d\theta_n\,d\Omega^n_{2-2\epsilon} dP_n^2\notag\\
&\times \(4\pi\)^{2\epsilon-3} \(P_n^2\)^{-\epsilon}\(1-\frac{P_{n-1}^2}{P_n^2}\)^{1-2\epsilon} \(\sin\theta_{n-1}\)^{1-2\epsilon} d\theta_{n-1}\,d\Omega^{n-1}_{2-2\epsilon}dP_{n-1}^2\notag\\
&\dots\notag\\
&\times \(4\pi\)^{2\epsilon-3} \(P_3^2\)^{-\epsilon}\(1-\frac{P_2^2}{P_3^2}\)^{1-2\epsilon} \(\sin\theta_2\)^{1-2\epsilon} d\theta_2\,d\Omega^2_{2-2\epsilon}dP_2^2 \notag\\
&\times 
\frac{\(2\pi\)^{2\epsilon-2}}{8}\frac{\(\pt^2\)^{-\epsilon}}{\sqrt{P_2^2}
  \sqrt{\frac{P_2^2}{4}\(1-\frac{M^2}{P_2^2}\)^2-\pt^2}
   } d\Omega^1_{2-2\epsilon} d\pt^2,
\end{align}
where, in  
view of the fact that we are interested in the transverse momentum
spectrum of the $(n+1)$-th particle $\mathcal{S}\(p\)$, we
have parametrized its phase space in terms of $\pt$  using
Eq.~(\ref{eq:phi22}), while 
the identification of $p_T$ with the transverse
momentum in the center-of-mass frame of the hadronic collision, as
well as the reason why 
all other phase spaces are parametrized in
terms of $\theta_i$ in  Eq.~(\ref{eq:phi21}), will be
clear shortly.

The domain of integration over $\pt^2$ in
Eq.~(\ref{eq:Phin1})
is
\begin{equation}\label{ptrange}
0\le \pt^2\le \frac{P_2^2}{4}\left(1-\frac{M^2}{P_2^2}\right)^2,
\end{equation}
so that the overall domain of integration
Eqs.~(\ref{eq:prange},\ref{eq:ptrange}) over the $n$ dimensional
variables which characterize the $n$ emissions  is ordered,
with the upper limit of integration of $p^2_T$ being set by $P_2^2$,
and then each of the integrations over $P_i^2$ with $i\ge 2$ being
limited from above by $P_{i+1}^2$, with the aforementioned
identification $P_{n+1}^2=\hat s$. Of course, in the collinear limit
$\pt\to0$ this leads to the standard strongly-ordered integration region
which leads to collinear factorization and  Altarelli-Parisi
evolution. This is then the form of the phase space which is most
convenient in order to discuss the total inclusive cross section and
specifically its soft limit, as was done in Refs.~\cite{Forte:2002ni,Bolzoni:2005xn}.

However, in order  to consider instead the case in which $\pt$ is
kept fixed,  the
integration region  can be re-expressed by taking
$\pt$ as outer integration variable. In this case the integration
range over $\pt$ is only
limited from above according to  Eq.~(\ref{eq:ptrange}), while the
integration over all $P_i^2$ is now in the range
\begin{equation}
\label{eq:newptrange}
\left(\sqrt{M^2+\pt^2}+ \pt\right)^2   \le P^2_i  \le P^2_{i+1}.
\end{equation}

The final  simplification of the  phase space Eq.~(\ref{eq:Phin1}) is
achieved by rewriting it in terms of the dimensionless variables
\begin{align}
z_i=\frac{P_i^2}{P_{i+1}^2}, 
\end{align}
along with $\tau^\prime$ Eq.~(\ref{tauprimedef}) and $\xi_p$ Eq.~(\ref{xidef}).
We thus arrive to our final expression for the phase space:
\begin{align}
\label{eq:Phin1mod2}
d\Phi&=   \tau^\prime\,\(4\pi\)^{2\epsilon-3} Q^{2-2\epsilon}
{\tau^\prime}^{\epsilon-1}\(1-z_n\)^{1-2\epsilon} \(\sin\theta_n\)^{1-2\epsilon}
d\theta_n\,d\Omega^n_{2-2\epsilon}\frac{dz_n}{z_n} \notag\\
&\times \(4\pi\)^{2\epsilon-3} Q^{2-2\epsilon} \(\frac{\tau^\prime}{z_n}\)^{\epsilon-1}\(1-z_{n-1}\)^{1-2\epsilon} \(\sin\theta_{n-1}\)^{1-2\epsilon} d\theta_{n-1}\,d\Omega^{n-1}_{2-2\epsilon}\frac{dz_{n-1}}{z_{n-1}}\notag\\
&\dots\\
&\times \, \(4\pi\)^{2\epsilon-3} Q^{2-2\epsilon} \(\frac{\tau^\prime}{z_n
  \dots z_3}\)^{\epsilon-1}\(1-z_2\)^{1-2\epsilon}
\(\sin\theta_2\)^{1-2\epsilon}
d\theta_2\,d\Omega^2_{2-2\epsilon}\frac{dz_2}{z_2} \notag\\
&\times \frac{\(2\pi\)^{2\epsilon-2}}{4}\(\frac{\xi_p}{\(\sqrt{1+\xi_p}+\sqrt{\xi_p}\)^2}\)^{-\epsilon}\notag\\
&\frac{
  Q^{2-2\epsilon} }{\sqrt{\(1-\frac{\tau^\prime}{z_n
      \dots
      z_2}\)\(1-\(\sqrt{1+\xi_p}-\sqrt{\xi_p}\)^4\frac{\tau^\prime}{z_n\dots
      z_2}\)}  }d\Omega^1_{2-2\epsilon} {d\xi_p}.\notag
\end{align}

For  fixed $\pt$, the  integration over the
set of $n-1$ dimensional variables $P_i^2$ with $2\le i\le n$ now
becomes the integral over the $n-1$ variables $z_i$, $2\le i\le
n$. Its range is
\begin{equation}
\label{eq:zintrange}
\frac{\tau^\prime}{z_n z_{n-1}\dots z_{i+1}} \le z_i\le 1 .
\end{equation}
We  note that the phase space Eq.~(\ref{eq:Phin1mod2}),
integrated over the range Eq.~(\ref{eq:zintrange}) has the structure
of a multiple convolution, and thus it factorizes upon taking a Mellin transform
with respect to $\tau^\prime$, with $n-1$ identical factors depending
on momenta $k_i$, $2\le i\le n$, and one factor depending on the
two-body phase space of the leading-order process in which a single
parton with momentum $\pt$ recoils against the heavy state  $\mathcal{S}\(p\)$.
When comparing to the phase-space  factorization of
Refs.~\cite{Forte:2002ni,Bolzoni:2005xn} it should be kept in mind
that Eq.~(\ref{eq:Phin1mod2})
holds at the differential level in $\pt$  because
$\tau^\prime$ Eq.~(\ref{tauprimedef}) is $\pt$ dependent. 

Note that this factorization ensues thanks to the choice of parametrizing 
momenta $k_i$ in terms of the polar angles $\theta_i$: had we chosen
to also parametrize them in terms of their transverse component, the
Jacobian factors would have spoiled the convolution structure. It is
important however to remember that this factorization has been
obtained thanks to the choice Eq.~(\ref{eq:phi21}-\ref{eq:phi22}) of writing each
two-body phase space in the respective center-of-mass frame. Now, in
the infrared limit in which the energy of all emitted
partons vanishes, all these reference frames coincide: this is the same
mechanism which underlies 
standard CFP factorization~\cite{Curci:1980uw}, and leads to
factorization in the eikonal limit~\cite{Forte:2002ni} if the
amplitude factorizes~\cite{Contopanagos:1996nh}. But for generic momenta,
this phase-space factorization  is not useful
because it only follows by choosing a different reference frame for
each emission.

In the threshold limit the squared amplitude has  infrared and
collinear singularities respectively proportional to $(1-z_i)^{-2}$
and $(\sin\theta_i)^{-2}$, which, when combined with the phase space
Eq.~(\ref{eq:Phin1mod2}), lead each to a simple pole in $\epsilon$
upon integration over the emitted particle's momenta, and are resummed
into the Sudakov and soft exponentials of
Eq.~(\ref{eq:largeNjoint}). Comparing the resummed result
Eqs.~(\ref{eq:largeNjoint}-\ref{eq:SN}) to the phase-space
Eq.~(\ref{eq:Phin1mod2}), and noting that the amplitude can only
depend polynomially on momenta, we see that the logs resummed into the
Sudakov exponent originate from the factor $(1-z)^{-2\epsilon}$
interfering with the $\epsilon$ poles due to the infrared and
collinear integrations over the amplitude. As explained in
Ref.~\cite{Forte:2002ni} the fact that the dependence is driven by a
phase-space factor reflects its origin from the upper limit of
integration on energy of the emitted parton in logarithmically
divergent integration.

Furthermore, the logarithmic $\xi_p$ dependent factor in the soft
anomalous dimension Eq.~(\ref{eq:SN}) is seen to originate from the
Jacobian factor
$\(\xi_p/\(\sqrt{1+\xi_p}+\sqrt{\xi_p}\)^2\)^{-\epsilon}$ related
to the leading-order process: it is thus  due to
interference between this large-angle radiation, and the $\epsilon$
poles due to soft emission.
It is interesting to observe that 
\begin{equation}\label{eq:noncomm1}
\lim_{\xi_p\to0}\( \sqrt{1+\xi_p}+\sqrt{\xi_p}\)=1,
\end{equation}
so logs coming from the soft
anomalous dimension Eq.~(\ref{eq:SN}) are absent in the small $\pt$
limit.

On the other hand,
in the same limit the Jacobian factor in the denominator of the
phase-space related to  the leading-order process in the last line of
Eq.~(\ref{eq:Phin1mod2}) becomes
\begin{equation}
\label{eq:extraIR}
\lim_{\xi_p\to0}\frac{1}{\sqrt{\(1-z\)\(1-\(\sqrt{1+\xi_p}-\sqrt{\xi_p}\)^4z\)}}=
\frac{1}{1-z},
\end{equation}
thus leading to extra infrared divergences. In fact
the resummed Mellin-space transverse momentum
distribution for finite $\pt$ in the large-$N$ limit vanishes as
$\frac{1}{\sqrt N}$. As mentioned in
the end of Sect.~\ref{subsec:threshold}, this is the correct large $N$
behaviour of the transverse momentum distribution to any perturbative
order. However,
if one were to take the $\pt \to0$ limit and
use the expression on the r.h.s. of Eq.~(\ref{eq:extraIR}) , one would
instead find that the 
transverse momentum distribution grows as $\ln N$; we will come back
to this point in Sect.~\ref{sec:phasespacelogb} below.
 This shows that
the soft and small $\pt$ limits do not commute, as one might expect on
physical grounds, given that for finite $p_T$ at least one parton must
recoil against the colorless final state. Therefore, resummed
expressions derived in the  $\pt\to0$ limit do not provide a fully
resummed result for  finite
$\pt$.

On the other hand, we also note that the threshold and small $\pt$
limits differ, in that the kinematic configurations which contribute
to the two limits are different. Indeed, in the threshold limit $s\to Q^2$ 
the only allowed radiation is that which leaves $Q^2$
Eq.~(\ref{scaldef}) unchanged. 
This corresponds to radiation of partons
which are either infrared, or collinear to the parton recoiling
against   $\mathcal{S}\(p\)$. However, the latter collinear radiation
does not lead to logarithmic enhancement because $\pt$ is large, and
thus does not contribute to the limit. On the other hand, 
in the small $\pt$ limit
one may also have radiation of partons whose transverse momenta are
collinear but not soft, in that they are subject to a large boost in
the longitudinal direction.
It follows that resummed expressions derived for finite $\pt$ do
not lead to the correct resummed small $\pt$ limit because not all
relevant kinematic configurations are included.

If one wishes to also include collinear non-soft radiation, the
phase-space factorization Eq.~(\ref{eq:Phin1mod2}) does not help
because, as mentioned, it only holds choosing a different (boosted)
reference frame for each emission. In such case a different phase
space factorization must be adopted, as we discuss in the next
section.

\subsection{Small $\pt$}
\label{sec:phasespacelogb}

In order to study the small $\pt$ limit we need a factorization of
phase space which holds even when longitudinal momenta are not 
small: this can be done by separating the longitudinal and transverse
momentum integrations. Our eventual task is to construct a
combined resummed result that reproduces threshold resummation of the
total cross section upon integration over transverse momentum. We have
seen in the previous section however that threshold resummation at
finite $\pt$ does not include collinear contributions, which do
contribute to the threshold limit at the integrated
level~\cite{Catani:1989ne,Sterman:1986aj}. We will strive
to achieve this phase-space factorization in such a way that the soft
limit at the integrated level is not spoiled.

We start from the general form for the phase-space for process 
Eq.~(\ref{eq:process}) in $d=4-2\epsilon$ dimensions:
\begin{align}
\label{eq:phasespacestart}
d\Phi_{n+1}\(p_1,p_2;p,k_1,\dots,k_n\)&=\frac{d^{3-2\epsilon}
  p}{\(2\pi\)^{3-2\epsilon} 2 \sqrt{M^2+|\vec p|^2}}\frac{d^{3-2\epsilon} k_1}{\(2\pi\)^{3-2\epsilon} 2 E_1}\dots \frac{d^{3-2\epsilon} k_n}{\(2\pi\)^{3-2\epsilon} 2 E_n}\notag\\
&\(2\pi\)^{4-2\epsilon} \delta^{(4-2\epsilon)}\(p_1+p_2-p-k_1-\dots-k_n\).
\end{align}
Representing the transverse momentum constraint as a Fourier transform
with respect to an impact parameter $\vec b$ conjugate to $\vec \pt$
we get
\begin{align}
\label{eq:phasespacesmallpt1}
&d\Phi_{n+1}\(p_1,p_2,p;k_1,\dots,k_n\)=\(2\pi\)^{2-2\epsilon}\int
d^{2-2\epsilon} b \frac{|\vec p| d|\vec p|\, \(\pt^2\)^{-\epsilon}
  d\pt^2 d\Omega_{2-2\epsilon} e^{i \vec{b} \cdot \vec{p}_{\sss
      T}}}{2\sqrt{M^2+|\vec p|^2}\(2\pi\)^{3-2\epsilon}\sqrt{|\vec p|^2-\pt^2}}\notag\\
&\frac{dE_1\, \(k_{\sss T_1}^2\)^{-\epsilon} dk_{\sss T_1}^2 d\Omega_{2-2\epsilon} e^{i \vec{b} \cdot \vec{k}_{\sss T_1}}}{4\(2\pi\)^{3-2\epsilon}\sqrt{E_1^2-k_{\sss T_1}^2}}\dots\frac{dE_n\, \(k_{\sss T_n}^2\)^{-\epsilon} dk_{\sss T_n}^2 d\Omega_{2-2\epsilon} e^{i \vec{b} \cdot \vec{k}_{\sss T_n}}}{4\(2\pi\)^{3-2\epsilon}\sqrt{E_n^2-k_{\sss T_n}^2}}\notag\\
&\delta\(\sqrt{\hat s}-\sqrt{M^2+|\vec p|^2}-E_1-\dots-E_n\)\delta\(p_z-k_{1_z}-\dots-k_{n_z}\),
\end{align}
where we have also traded the integral over the longitudinal momentum
component for an integral over energy.

In order to separate the transverse  and longitudinal momentum
dependence,  it is convenient to adopt a Sudakov parametrization:
\label{eq:Sudakovpar}
\begin{align}
k_1&=\(1-z_1\)p_1-\frac{k_{\sss T_1}^2}{\(1-z_1\) s}p_2+k_{\sss T_1}\\
k_2&=z_1\(1-z_2\)p_1-\frac{k_{\sss T_2}^2}{z_1\(1-z_2\)s}p_2+k_{\sss T_2}\\
&\dots\notag\\
k_n&=z_1\dots z_{n-1}\(1-z_n\)p_1-\frac{k_{\sss T_n}^2}{z_1\dots z_{n-1}\(1-z_n\)s}p_2+k_{\sss T_n}.
\end{align}
In the small $\pt$ limit we then get
\begin{align}
\label{eq:phasespacesmallpt2}
&d\Phi_{n+1}\(p_1,p_2;p,k_1,\dots,k_n\)=\(2\pi\)^{2-2\epsilon}\int d^{2-2\epsilon} b\frac{ d|\vec p| \(\pt^2\)^{-\epsilon} d\pt^2 d\Omega_{2-2\epsilon} e^{i \vec{b} \cdot \vec{p}_{\sss T}}}{4\sqrt{M^2+|\vec p|^2}\(2\pi\)^{3-2\epsilon}}\notag\\
&\frac{dz_1\, \(k_{\sss T_1}^2\)^{-\epsilon} dk_{\sss T_1}^2 d\Omega_{2-2\epsilon} e^{i \vec{b} \cdot \vec{k}_{\sss T_1}}}{4\(2\pi\)^{3-2\epsilon}\sqrt{\(1-z_1\)^2-\frac{4}{s}k_{\sss T_1}^2}}\dots\frac{dz_n\, \(k_{\sss T_n}^2\)^{-\epsilon} dk_{\sss T_n}^2 d\Omega_{2-2\epsilon} e^{i \vec{b} \cdot \vec{k}_{\sss T_n}}}{4\(2\pi\)^{3-2\epsilon}\sqrt{\(1-z_n\)^2-\frac{4}{s z_1^2\dots z_{n-1}^2} k_{\sss T_n}^2}}\notag\\
&\delta\(\sqrt{\hat s}-\sqrt{M^2+|\vec p|}-|\vec p|\)\delta\(|\vec p|-\(1-z_1\dots z_n\)\frac{\sqrt{s}}{2}\)+\Ord\(\frac{1}{b}\)
\end{align}
where   we have changed integration variable from $k_i$ to $z_i$ and
we have denoted by $\Ord\(\frac{1}{b}\)$ neglected terms which lead to power
suppressed contributions in the small $p_T$ limit. Note
that we have kept the $k_{\sss T_i}^2$ dependence in the
Jacobian square-root factors, even though it is also
$\Ord\(\frac{1}{b}\)$, for reasons to be discussed shortly.

The final expression for the phase space is obtained by introducing
dimensionless variables $x$, Eq.~(\ref{eq:partx}) and
$\xi_i=\frac{k_{\sss T,i}^2}{M^2}$ defined in analogy to $\xi_p$,
Eq.~(\ref{xidef}), and 
performing the angular integrations
\begin{align}
\label{eq:angularint}
\int d\Omega_{2-2\epsilon} e^{i \vec{b} \cdot \vec{k_{\sss T}}}=\(b k_{\sss T}\)^\epsilon \(2\pi\)^{1-\epsilon} J_{-\epsilon}\(b k_{\sss T}\),
\end{align}
where $J_{-\epsilon}$ is implicitly defined by
Eq.~(\ref{eq:angularint}) and it reduces to the Bessel function $J_0$
when $\epsilon\to0$.
We get
\begin{align}
\label{eq:phasespacesmallpt3}
d\Phi_{n+1}(p_1,p_2&;q,k_1,\dots,k_n)
=x \frac{\pi^{3-2\epsilon}}{\Gamma\(1-\epsilon\)}d\xi_p\int db^2\,
\(b \pt\)^{-\epsilon}b^{n\epsilon}
 J_{-\epsilon}\(b \pt\) \notag\\
&J_{-\epsilon}\(b k_{\sss T_1}\)\frac{ M^{-\epsilon} \(\xi_1\)^{-\frac{\epsilon}{2}}dz_1d\xi_1}{4\(2\pi\)^{2-\epsilon}\sqrt{\(1-z_1\)^2-4 x\xi_1}}\notag\\\dots\notag\\
&J_{-\epsilon}\(b k_{\sss T_n}\)\frac{M^{-\epsilon} \(\xi_n\)^{-\frac{\epsilon}{2}}dz_nd\xi_n}{4\(2\pi\)^{2-\epsilon}\sqrt{\(1-z_n\)^2-\frac{4x}{ z_1^2\dots z_{n-1}^2} \xi_n}}\notag\\
&\times\delta\(x-z_1\dots z_n\)+\Ord\(\frac{1}{b}\) .
\end{align}
The integration range over transverse momenta is
\begin{equation}\label{trmomrange}
0 \le \xi_i \le \frac{z_1^2\dots z_{i-1}^2\(1-z_i\)^2}{4x},
\end{equation}
while all $z_i$ range from $0\le z_i\le 1$. 

The expression of the phase space Eq.~(\ref{eq:phasespacesmallpt3})
would have the structure of a convolution, and thus factorize upon
Mellin transformation with respect to $x$, were it not for
the $\xi_i$ terms in the denominator. Up to $\Ord\(\frac{1}{b}\)$
corrections, these can be simplified by letting all
$\xi_i\to0$. This then leads to a factorized form of phase space which,
when combined with a suitably factorized and renormalization-group
improved form of the amplitude, leads
to transverse momentum resummation
Eq.~(\ref{eq:smallptjoint3}). However, this also leads to a result
which in the soft limit does not have have the correct behaviour and
is beset by a spurious logarithmic rise, because of the
non-commutativity of limits which we have already seen in
Eq.~(\ref{eq:extraIR}) when discussing the factorization of phase
space in the soft limit. 
We will now show first, how transverse momentum resummation is usually
derived
in the $\xi_i\to0$ limit,  then, why this expansion spoils
the soft limit at the integrated level, and finally, how factorization
of phase space can be obtained while preserving the correct soft
limit. 

The
$ \xi_i\to0$  limit of  Eq.~(\ref{eq:phasespacesmallpt3}) must be taken
with some care because new infrared singularities arise in the
limit. These can be taken care of by rewriting the Jacobian
square-root factors as
\begin{align}\label{eq:jacobexp}
\frac{1}{\sqrt{(1-z_1)^2-4x \xi_n}}&=\[\frac{1}{\sqrt{\(1-z_1\)^2-4x
  }}\]_+^{z_1}\notag\\
&+ \frac{1}{2}\ln\frac{1-\sqrt{1- 4x \xi_1}}{1+\sqrt{1-4 x \xi_1}}
    \delta\[z_1-\(1-2\sqrt{x \xi_n}\)\],
\end{align}
and similarly for all $z_i$,
where the plus distribution is defined as
\begin{equation}
\label{eq:plusx}
\int_0^1 dz\, g\(z\) \plus{f\(z\)}^{z}=\int_0^1
dz\,\[g\(z\)-g\(1\)\]f\(z\).
\end{equation}
Substituting Eq.~(\ref{eq:jacobexp})
in Eq.~(\ref{eq:phasespacesmallpt3}) we can now safely
take the $\xi_i\to0$ limit, with the result 
\begin{align}
\label{eq:phasespacesmallpt4}
d\Phi_{n+1}(p_1,p_2;&q,k_1,\dots,k_n)=x \frac{\pi^{3-2\epsilon}}
{\Gamma\(1-\epsilon\)}d\xi_p\int db^2\,\(b \pt\)^{-\epsilon} b^{n\epsilon}
J_{-\epsilon}\(b \pt\)
\notag\\
&J_{-\epsilon}\(b k_{\sss T_1}\)\frac{ M^{-\epsilon}
  \(\xi_1\)^{-\frac{\epsilon}{2}}}{4\(2\pi\)^{2-\epsilon}}
\left[\plus{\frac{1}{1-z_1}}-\delta\(1-z_1\)\frac{1}{2}\ln\xi_1\right]
dz_1 d\xi_1
\notag\\
&\dots\\
&\,J_{-\epsilon}\(b k_{\sss T_n}\)\frac{M^{-\epsilon}
  \(\xi_n\)^{-\frac{\epsilon}{2}}}{4\(2\pi\)^{2-\epsilon}}\left[\plus{\frac{1}{1-z_n}}-\delta\(1-z_n\)\frac{1}{2}\ln\xi_n\right]dz_nd\xi_n
\notag\\
&\qquad\qquad \times\delta\(x-z_1\dots z_n\)+\Ord\(\frac{1}{b}\),\notag
\end{align}
where now the integration range over transverse momenta is
\begin{equation}\label{trmomrangecss}
0 \le \xi_i \le \infty 
\end{equation}
for all $\xi_i$

The phase space Eq.~(\ref{eq:phasespacesmallpt4}) factorizes upon
Mellin transform. It can be used to derive the
standard transverse momentum resummation  Eq.~(\ref{eq:sudpt}): 
in the small $\pt$ limit the squared amplitude 
has collinear singularities (both soft and non-soft) 
 which are
resummed into the Sudakov form factor Eq.~(\ref{eq:sudpt}), with the
$A^{\pt}(\alpha_s)$ term driven by the interference with the infrared
singularity of the DGLAP anomalous dimension with the
contributions to the phase space Eq.~(\ref{eq:phasespacesmallpt4})
enhanced by $\ln\xi_i$.

However, this resummed expression, Eq.~(\ref{eq:smallptjoint3}),
 does not reproduce
the correct behaviour of the total cross section in the soft limit
upon integration over $\pt$. This can be seen by noting that,
integrating
over the phase space Eq.~(\ref{eq:phasespacesmallpt3}) in the range
Eq.~(\ref{trmomrange}) and expanding in powers of $\xi_i$,  one ends up with  integrals (one for each
emission) of the form
\begin{align}\label{eq:unstexp}
\int_0^{\frac{\(1-z\)^2}{4}} d\xi\, \frac{1}{\sqrt{\(1-z\)^2-4 \xi}}&=\frac{1}{1-z}\int_0^{\frac{\(1-z\)^2}{4}} d\xi 1+\frac{2 \xi}{\(1-z\)^2}+\frac{6\xi^2}{\(1-z\)^4}+\dots\notag\\
&=\frac{\(1-z\)}{4}\(1+\frac{1}{4}+\frac{1}{8}+\dots\),
\end{align}
where for definiteness we have considered 
integration with respect to $\xi_1$. It is apparent that all terms in the expansion in powers
of $\xi_i$ are in fact of the same order after integration. Hence,
only retaining the first term in this expansion as it was done in the
derivation of Eq.~(\ref{eq:phasespacesmallpt4}) spoils the $x\to1$
limit at the integrated level. 

The relevant power counting is clear in Fourier-Mellin space, 
where resummed expressions are derived, and in which the expansion
  Eq.~\eqref{eq:unstexp} becomes a series in powers of $\frac{1}{b}$:
\beq
\label{eq:unstexp2}
\int_0^1 dz\,z^{N-1} \int_0^{\frac{\(1-z\)^2}{4}}d\xi\,J_0\(b M \sqrt{\xi}\)\frac{1}{\sqrt{\(1-z\)^2-4 \xi}}=\frac{2}{b^2 M^2}\(1-\frac{4 N^2}{b^2 M^2}+\frac{16 N^4}{b^4 M^4}+\dots\).
\eeq
Namely, it is clear that the expansion parameter is
$\frac{N}{b}$. Because the inclusive cross section is obtained by
taking $b=0$, truncating this expansion spoils the large $N$ behaviour
at the inclusive level. Hence, in order to preserve the threshold
limit at the inclusive level, transverse momentum resummation must be
performed by taking the limit $b\to\infty$ at fixed $\frac{N}{b}$, rather
than at fixed $N$.

A form of the phase space which does factorize upon Mellin transform,
but which does not spoil the soft limit at the integrated level can be
obtained by evaluating the phase space
Eq.~\eqref{eq:phasespacesmallpt3} and integration range
Eq.~(\ref{trmomrange}) in the soft limit, namely
\begin{align}
\label{eq:phasespacesmallpt5}
d\Phi_{n+1}(p_1,p_2&;q,k_1,\dots,k_n) =x \frac{\pi^{3-2\epsilon}}{\Gamma\(1-\epsilon\)}d\xi_p\int db^2\,\(b \pt\)^{-\epsilon}b^{n\epsilon}
J_{-\epsilon}\(b \pt\) \notag\\
&J_{-\epsilon}\(b k_{\sss T_1}\)\frac{ M^{-\epsilon} \(\xi_1\)^{-\frac{\epsilon}{2}}dz_1d\xi_1}{4\(2\pi\)^{2-\epsilon}\sqrt{\(1-z_1\)^2-4 z_1\xi_1}}\notag\\\dots\notag\\
&J_{-\epsilon}\(b k_{\sss T_n}\)\frac{M^{-\epsilon} \(\xi_n\)^{-\frac{\epsilon}{2}}dz_nd\xi_n}{4\(2\pi\)^{2-\epsilon}\sqrt{\(1-z_n\)^2-4z_n \xi_n}}\notag\\
&\times\delta\(x-z_1\dots z_n\)+\Ord\(\frac{1}{b}\)+\Ord\(\frac{1}{N}\)
\end{align}
and
\begin{equation}\label{trmomrangejoint}
0 \le \xi_i \le \frac{\(1-z_i\)^2}{4z_i},
\end{equation}
where we have denoted by $\Ord\(\frac{1}{N}\)$ terms which do not
contribute to the threshold limit. In going from
Eq.~(\ref{eq:phasespacesmallpt3}) to
Eq.~(\ref{eq:phasespacesmallpt5}) we have retained the  unexpanded
phase-space factor of Eq.~(\ref{eq:unstexp2}), but in it we have
neglected the factors of $\frac{1}{z_i^2}$ which multiply $\xi$, which
would spoil the convolution structure, but correct it by
$\Ord\(\frac{1}{N}\)$ terms. As a consequence, the result
now has the form of a
convolution, and  it factorizes upon Mellin transform when
combined with an amplitude that also has the structure of a
convolution, but now it also
includes all the contributions in the phase space at large $b$ of
$\Ord\(\frac{N^k}{b^k}\)$.  

Recalling that transverse momentum distributions are
convolutive with respect to the partonic scaling variable $x$
Eq.~(\ref{eq:partx}), one has
\begin{align}
\label{eq:phasespacesmallpt6}
&\int_0^1d x x^{N-1}\int d\xi_1d\xi_2\dots d\xi_n
\frac{d\Phi_{n+1}\(p_1,p_2;q,k_1,\dots,k_n\)}{d\xi_p}=\notag\\
&=\frac{\pi^{2-\epsilon} b^{n\epsilon}}{\Gamma\(1-\epsilon\)}\int db^2\,\(b \pt\)^{-\epsilon} J_{-\epsilon}\(b \pt\) \left(\sqrt{1+\xi_p}-\sqrt{\xi_p}\right)^{-2N}\notag\\
&\Bigg[\int_0^\infty d\xi\,\(\sqrt{1+\xi}-\sqrt{\xi}\)^{2N}J_{-\epsilon}\(b k_{\sss T}\)M^{-\epsilon} \(\xi\)^{-\frac{\epsilon}{2}}\notag\\
&\int_0^1 dz\,z^{N-1}\frac{1}{4\(2\pi\)^{2-\epsilon}\sqrt{\(1-z\)\(1-\(\sqrt{1+\xi}-\sqrt{\xi}\)^4 z\)}}\Bigg]^n+ \Ord\(\frac{1}{b}\)+\Ord\(\frac{1}{N}\),
\end{align}
where we have used Eq.~(\ref{eq:partx}), and we have interchanged the $\xi_i$ and $z_i$ integrations and
performed the change of integration variables $z_i\to
z'_i\(\sqrt{1+\xi_i}-\sqrt{\xi_i}\)^2$. 

It is interesting to observe that, when combined with a matrix element
which behaves as a constant when $z_i\to1$, the phase space
Eq.~(\ref{eq:phasespacesmallpt6}) leads to
the correct large-$N$ behaviour of the leading-order  transverse
momentum distribution, which, as discussed in
Sect.~\ref{subsec:threshold}, is
$\frac{d\hat{\sigma}_{ij}^{\rm LO}}{d\xi_p}\(N,\xi_p\)\toinf{N}
\frac{1}{\sqrt{N}}$. Indeed, at leading order, $n=1$ and $\xi_p=\xi$ in Eq.~(\ref{eq:phasespacesmallpt6}), so the Mellin transform of
  the transverse momentum distribution behaves as
\small
\begin{align}\label{eq:largenok}
\int_0^1 dz\,z^{N-1}\frac{1}{\sqrt{\(1-z\)\(1-\(\sqrt{1+\xi_p}-\sqrt{\xi_p}\)^4 z\)}}&=
\frac{\sqrt{\pi}}{2\sqrt{N}}\(\(\frac{\xi_p}{1+\xi_p}\)^{\frac{1}{4}}+\(\frac{1+\xi_p}{\xi_p}\)^{\frac{1}{4}}\)\notag\\
&+\Ord\(\frac{1}{N}\).
\end{align}
\normalsize
Had the form Eq.~(\ref{eq:phasespacesmallpt4}) of the phase space been
used instead, one would take $\xi_1=0$ first, and find a Mellin-space
transverse momentum distribution which displays a spurious
logarithmic growth
\beq\label{eq:spurious}
\int_0^1 dz\,z^{N-1}\plus{\frac{1}{1-z}}=-\ln N +\Ord\(1\).
\eeq
This is another manifestation of the fact that truncation of the
expansion Eq.~(\ref{eq:unstexp}) spoils the soft limit, because higher
order terms are enhanced by powers of $1-x$ despite being suppressed
by powers of $\xi_i$

\section{Resummation}
\label{sec:joint}

Using the phase-space arguments of Sect.~\ref{sec:phasespace} we can
now construct a combined resummed expression which satisfies all of the
requirements discussed in the introduction. First, in Sect.~\ref{subsec:smallptjoint} we will present a
formal construction of our resummed formula and argue that it has the
required properties, then, in Sect.~\ref{subsec:Higgsjoint} we will
provide explicit expressions in the case of Higgs production in gluon
fusion up to NNLL, and check explicitly that in the three limits of
large $N$ at fixed $\pt$, small $\pt$ for generic $N$, and large $N$
at the integrated level, known NNLL results are reproduced. This will
provide us with nontrivial insight on the relation between these
different resummations. 

Here and henceforth will refer to the
$N\to\infty$ or $x\to1$ limit as threshold limit. Resummation in this
limit, which we will refer to as 
threshold resummation, includes, to a given logarithmic order and to all
orders in $\alpha_s$, powers of $\ln N$ or $\ln
(1-x)$, up to contributions which have a relative power suppression
in $1/N$ or $1-x$. Unless otherwise stated, threshold
resummation will refer to the resummation being performed at fixed and
finite $\pt$.
We will refer to the $b\to\infty$ or $\pt \to 0$
limit as small $\pt$ limit.
Resummation in this
limit, which we will refer to as 
transverse momentum  resummation, includes, to a given logarithmic
order, and to all
orders in $\alpha_s$, powers of $\ln b$ or $\ln \xi_p$, up to contributions which have a
relative power suppression 
in $1/b$ or $\xi_p$.

\subsection{The combined resummed result}
\label{subsec:smallptjoint}
The combined resummed result can be constructed based on two sets of
observations, which substantially rely on the phase-space analysis of Sect.~\ref{sec:phasespace}. 

The first concerns the relation between
transverse momentum resummation and threshold resummation at finite $\pt$.
 The key observation is
that the threshold limit and the small $\pt$ limit  do not
commute, Eqs.~(\ref{eq:noncomm1}-\ref{eq:extraIR}). It then follows
that because of  Eq.~(\ref{eq:extraIR}) [see
also Eqs.~(\ref{eq:largenok}-\ref{eq:spurious})], results
derived in the small
$\pt$ limit display an unphysical growth with $N$ in the threshold
limit. However, this unphysical growth can be removed, and the correct
large $N$ behaviour for finite $\pt$ can be restored, if the
form Eq.~(\ref{eq:phasespacesmallpt6}) 
of phase-space is used in the derivation of
transverse momentum resummation, instead of the usual one
Eq.~(\ref{eq:phasespacesmallpt4}). 
This form of the phase space is compatible with
factorization leading to transverse momentum resummation, and differs
from the standard result by terms which are power-suppressed in the small
$\pt$ limit: it thus leads to an alternative form of transverse
momentum resummation which differs from the usual one by
power-suppressed terms $\Ord\(\frac{1}{b}\)$ at fixed $N$. 

However, even after doing this, because
 of Eq.~(\ref{eq:noncomm1}), results derived in the small
$\pt$ limit do not contain all of the large $N$ logs which are present 
at finite $\pt$.
Furthermore, while  soft radiation (which is responsible for
logs in the
threshold limit)  is also collinear, collinear
radiation (which is responsible for
logs in the
small $\pt$  limit)  can be non-soft: as a consequence, transverse momentum
resummation includes logarithmic contributions that do not contribute
to the soft limit. 

The upshot is that transverse momentum resummation in the soft limit
displays an unphysical growth with $N$ which may be removed by
inclusion of terms which are power-suppressed in the small $\pt$
limit, but even having done that, transverse momentum resummation in
the large $N$ limit still does not include all
large logs which contribute to the threshold limit of the full
cross section and are included in threshold resummation. Conversely,
threshold resummation in the small $\pt$ limit does not include all
large logs which contribute to the small $\pt$ limit of the full
cross section and are included in transverse momentum resummation.

The second set of observations concerns the relation between
resummation of the transverse momentum distribution, and the threshold
limit of the inclusive cross section after integration over transverse
momentum. Here the key observation is that in order to be enhanced in the
threshold limit after integration over transverse momentum, a
contribution to the amplitude must have the highest allowed
power growth as $\pt \to 0$, otherwise integration over transverse
momentum  leads to a contribution which is power-suppressed in the
soft limit. It then follows that the full threshold limit at the
inclusive level can be obtained from integration over $\pt$ 
of all terms which
contribute to transverse momentum resummation. However, transverse
momentum resummation must be performed in the soft limit, i.e., 
retaining in the resummed expressions all terms which are enhanced in
the threshold limit even though they might be suppressed in the small
$\pt$ limit. In Fourier-Mellin space this  means that we must retain
the leading-order contribution in the
 large $b$ limit at $\frac{N}{b}$ fixed, rather than at fixed $N$  as
 it is usually done. 
This specifically implies that we must use
 the form Eq.~(\ref{eq:phasespacesmallpt6}) 
of phase-space, instead of the usual one
Eq.~(\ref{eq:phasespacesmallpt4}), because of
Eqs.~(\ref{eq:unstexp},\ref{eq:unstexp2}).

Based on this observation, a combined resummed
expression which satisfies all requirements can be constructed in two
steps. First, we construct a modified version 
$\frac{d\hat{\sigma}_{ij}^{\rm tr'}}{d\xi_p}$ 
of transverse momentum resummation
$\frac{d\hat{\sigma}_{ij}^{\rm tr}}{d\xi_p}$ 
Eq.~(\ref{eq:smallptjoint}), in which transverse
momentum resummation is performed in the soft limit, i.e. expanding in
powers of $\frac{1}{b}$ at fixed $\frac{N}{b}$ as just explained. 
This is obtained by noting that the squared amplitude for
real emission 
behaves as $|A|^2\tozero{\xi_i} \xi_i^{-1}$ as $\xi_i\to0$, which leads to the
infrared and collinear singularities resummed in
Eq.~(\ref{eq:smallptjoint3}), with infrared singularities canceled by
virtual terms.  Unlike the phase-space measure
Eq.~(\ref{eq:phasespacesmallpt3}), the square amplitude does not display
further infrared singularities as $x\to 1$ because standard power
counting arguments ensure that all infrared singularities in
propagators arise in the collinear if $\pt \to0$ limit. 
Hence the only $\frac{N}{b}$ terms arise due to
the phase space, and the   amplitude can be safely expanded in
powers of $\xi_p$ at fixed $N$. It is then sufficient to modify the
standard 
transverse momentum resummation Eq.~(\ref{eq:smallptjoint3}) in order
to account for 
the use of the phase space
Eqs.~(\ref{eq:phasespacesmallpt5}-\ref{trmomrangejoint}) instead of
the standard phase space
Eqs.~(\ref{eq:phasespacesmallpt4}-\ref{trmomrange})). 

Using the
Mellin-space expression Eq.~(\ref{eq:phasespacesmallpt6}) we then get
\begin{align}
\label{eq:smallptjoint}
&\frac{d\hat{\sigma}_{ij}^{\rm
    tr'}}{d\xi_p}\(N,\xi_p,\as\(M^2\),M^2\)=\sigma_0
\int_0^{\infty} db
\frac{b}{2} J_0\(b M
\sqrt{\xi_p}\)\(\sqrt{1+\xi_p}-\sqrt{\xi_p}\)^{-2N}\notag\\ &
\mathcal{H}_{ij}\(N,\as\(M^2\)\)\exp\left[\int_0^{\infty}d\xi 
  \(\sqrt{1+\xi}-\sqrt{\xi}\)^{2N} J_0\(b M
  \sqrt{\xi}\)\plus{\frac{\mathcal{B}\(N,\as\(M^2
      \xi\)\)}{\xi}}^{\pt}+\Ord\(\frac{1}{b}\)\right]\notag\\ 
&\exp\Bigg[\int_0^{\infty} d\xi \(\sqrt{1+\xi}-\sqrt{\xi}\)^{2N}
  J_0\(b M \sqrt{\xi}\)\int_0^1 dz\,z^{N-1}\notag\\ 
&\qquad \Bigg(\plus{\frac{2 A^{\pt}\(\as\(M^2 \xi\)\)}{\xi}}^{\pt} 
\plus{\frac{1}{\sqrt{\(1-z\)\(1-\(\sqrt{1+\xi}-\sqrt{\xi}\)^4 z\)}}}^{z}\notag \\
&\qquad+\delta\(1-z\)\frac{1}{2\(\sqrt{1+\xi}-\sqrt{\xi}\)^2}\notag\\
&\qquad\qquad\(2
  A^{\pt}\(\as\(M^2\xi\)\)\frac{\ln\(1+\xi\)}{\xi}-\plus{\frac{2
      A^{\pt}\(\as\(M^2\xi\)\)\ln\xi}{\xi}}^{\pt}\)\Bigg)+\Ord\(\frac{1}{b}\)\Bigg].
\end{align}
Here we have defined a second plus distribution
\begin{equation}\label{eq:pluspt}
\int_0^1 d\xi\,g\(\xi\) \plus{f\(\xi\)}^{\pt}=\int_0^1
  d\xi\,\[g\(\xi\)-g\(0\)\] f\(\xi\)\end{equation}
along with the usual one Eq.~(\ref{eq:plusx}); note that the
integration range in Eq.~(\ref{eq:pluspt}) is not the same as the
integration range in Eq.~(\ref{eq:smallptjoint}).
In order to get from
Eqs.~(\ref{eq:smallptjoint3},\ref{eq:phasespacesmallpt6}) to
Eq.~(\ref{eq:smallptjoint}) we have used the identity 
\begin{align}\label{eq:plusid}
&\frac{1}{\sqrt{\(1-z\)\(1-\(\sqrt{1+\xi}-\sqrt{\xi}\)^4
    z\)}}=\plus{\frac{1}{\sqrt{\(1-z\)\(1-\(\sqrt{1+\xi}-\sqrt{\xi}\)^4
      z\)}}}^{z}\nonumber\\
&\qquad +\delta(1-z)\frac{1}{2\(\sqrt{1+\xi}-\sqrt{\xi}\)^2}\left[\ln(1+\xi)-\ln\xi\right],
\end{align}
and we have further assumed cancellation of infrared singularities, so
all explicit $\frac{1}{\xi}$ terms are replaced by the corresponding
plus distributions. 

The notation $\Ord\(\frac{1}{b}\)$ in Eq.~(\ref{eq:smallptjoint})
means that all quantities are evaluated at leading
order in an expansion in powers of $b$ for fixed $\frac{N}{b}$, as
discussed above.  The  function 
 $A^{\pt}$ is then the  
same as in the standard transverse
momentum resummation Eq.~(\ref{eq:smallptjoint3}), 
while $\mathcal{B}$, which contains also contributions not enhanced as
$N\to\infty$, has to be determined by matching to the function $B$ of
Eq.~(\ref{eq:smallptjoint3}). Clearly, just like 
the function $B$ of Eq.~(\ref{eq:smallptjoint3}), also $\mathcal{B}$ is a matrix on the flavour space and the first exponential of Eq.~\eqref{eq:smallptjoint} must be viewed as path-ordered. 
Finally,  the function $\mathcal{H}_{ij}\(N,\as\)$ is determined by
first, matching it to a fixed order calculation, just like the
function $H_{ij}\(N,\as\)$ of
Eq.~(\ref{eq:smallptjoint3}) was (or, equivalently, matching
$\mathcal{H}_{ij}\(N,\as\)$ to  $H_{ij}\(N,\as\)$) and then, by observing
that in order to fully reproduce threshold resummation at the
integrated level, this function must be resummed to all orders in
$\alpha_s$ in the threshold limit up to the desired logarithmic order 
in $\ln N$, rather than just computed to finite order. Explicitly, we
let
\begin{align}
\label{eq:Hres}
\mathcal{H}_{ij}\(N,\as\(M^2\)\)= \mathcal{H}^{\rm f.o.}_{ij}\(N,\as\(M^2\)\)+\mathcal{H}^{0}_{ij}\(\as\(M^2\)\)
\exp\left[-D^{\pt}\(\as\(M^2\)\)\ln N\right].
\end{align}

The
expression Eq.~(\ref{eq:smallptjoint})  automatically leads to threshold resummation of the cross section
upon integration over transverse momentum, and it reproduces the
correct physical behaviour in the soft limit for finite $\pt$. However,
it does not contain all logarithmically enhanced terms in the soft
limit for finite $\pt$. In order to achieve
full threshold resummation for finite $\pt$, we combine
$\frac{d\hat{\sigma}_{ij}^{\rm tr'}}{d\xi_p}$
with the standard
threshold resummation 
$\frac{d\hat{\sigma}_{ij}^{\rm th}}{d\xi_p}$
Eq.~(\ref{eq:largeNjoint}) in a way which avoids double counting.
This can be  done by introducing  a matching function $T\(N,\xi_p\)$ such
that 
\begin{align}\label{eq:tlargen}
& \lim_{N \to \infty} T\(N,\xi_p\)= 1 \\ \label{eq:tsmallxi}
&\lim_{\xi_p \to 0} T\(N,\xi_p\)= 0,
\end{align} 
and letting
 \begin{align}
\label{eq:jointfinal}
\frac{d\hat{\sigma}_{ij}}{d\xi_p}\(N,\xi_p,\as\(\mu_R^2\),\mu_F^2\)&=\(1-T\(N,\xi_p\)\)\,\frac{d\hat{\sigma}_{ij}^{\rm
    tr'}}{d\xi_p}\(N,\xi_p,\as\(\mu_R^2\),\mu_F^2\)\notag \\
&\quad +T\(N,\xi_p\) \frac{d\hat{\sigma}_{ij}^{\rm
    th}}{d\xi_p}\(N,\xi_p,\as\(\mu_R^2\),\mu_F^2\).
\end{align}
Clearly, this avoids double-counting of any term in common between $\frac{d\hat{\sigma}_{ij}^{\rm
    tr'}}{d\xi_p}$ and  $\frac{d\hat{\sigma}_{ij}^{\rm
    th}}{d\xi_p}$. Furthermore, because of the limits
Eqs.~(\ref{eq:tlargen}-\ref{eq:tsmallxi}), the combined result
Eq.~(\ref{eq:jointfinal}) reproduces threshold resummation for large
$N$ and transverse momentum resummation  for small $\xi_p$, up to
corrections whose size can be tuned by choosing the form of the
matching function.
For instance choosing
\beq\label{eq:texplicit}
T\(N,\xi_p\)=\frac{N^k \xi_p^m}{1+N^k\xi_p^m}.
\eeq
the combined result differs from $\frac{d\hat{\sigma}_{ij}^{\rm
    tr'}}{d\xi_p}$ by $\Ord(\xi_p^m)$ corrections when $\xi_p\to0$,
and  from $\frac{d\hat{\sigma}_{ij}^{\rm
    th}}{d\xi_p}$ by $\Ord\left(\frac{1}{N^k}\right)$ corrections when
$N\to\infty$. 

Note that transverse momentum resummation
Eq.~(\ref{eq:smallptjoint}) is most easily performed by choosing
$\mu_R^2=\mu_F^2=M^2$, and  threshold resummation
Eq.~(\ref{eq:largeNjoint})  by choosing
$\mu_R^2=\mu_F^2=Q^2$ as given by Eq.~(\ref{scaldef}); however the
combined expression Eq.~(\ref{eq:jointfinal}) must be written with a
common choice of renormalization and factorization scale after
evolving either or both of its two terms to a common scale using
standard renormalization group expressions.

Equation~\eqref{eq:jointfinal} reproduces by
construction threshold 
resummation $\frac{d\hat{\sigma}_{ij}^{\rm
    th}}{d\xi_p}$ Eq.~(\ref{eq:largeNjoint}) in the large-$N$ limit at fixed
$\xi_p$, and transverse 
momentum resummation $\frac{d\hat{\sigma}_{ij}^{\rm
    tr'}}{d\xi_p}$ Eq.~(\ref{eq:smallptjoint}) in the $\pt\to 0$ limit
at fixed $N$, 
 because of Eq.~(\ref{eq:tlargen}) satisfied by the
matching function $T\(N,\xi_p\)$. Furthermore, the form of the
matching function also ensures that the threshold resummation
contribution $\frac{d\hat{\sigma}_{ij}^{\rm
    th}}{d\xi_p}$ does not contribute to the threshold limit of the
total cross section, because in the $\xi_p\to0$ limit it is regular
and thus upon integration it is free of logarithmic enhancement. On
the other hand, the small-$\xi_p$ logarithmic singularities of
Eq.~(\ref{eq:smallptjoint})  coincide to all logarithmic orders with
those of the transverse momentum resummation contribution $\frac{d\hat{\sigma}_{ij}^{\rm
    tr'}}{d\xi_p}$. Hence, upon integration over transverse momentum,
the combined result $\frac{d\hat{\sigma}_{ij}}{d\xi_p}$ has a threshold limit which coincides with that
of the integral of $\frac{d\hat{\sigma}_{ij}^{\rm
    tr'}}{d\xi_p}$, and thus it reproduces the threshold limit of the
total cross section for the reasons explained.

We conclude that Eq.~(\ref{eq:jointfinal}) satisfies all requirements
spelled out in the 
introductory Section~\ref{sec:introduction}, and thus we will take it
as our  final combined resummed
expression. In the next section we will work it out in a fully
explicit way for the case of Higgs production in gluon fusion, and
verify that indeed it has all its desired properties. In the course of
this argument, we will work out a more manageable form of our resummed
result.

\subsection{Higgs production}
\label{subsec:Higgsjoint}

We will now work out explicitly our combined resummed expression
Eqs.~(\ref{eq:jointfinal}) for the case of Higgs production in gluon
fusion up to NNLL accuracy.  This will allow us to explicitly check
that indeed our result has its desired properties. In the process of
doing this, we
will derive a simpler,  closed-form expression for our resummed result
by suitably neglecting subleading terms. 
Our starting point is
the expression of threshold resummation
$\frac{d\hat{\sigma}_{ij}^{\rm th}}{d\xi_p}$
Eq.~(\ref{eq:largeNjoint}) and transverse momentum resummation
$\frac{d\hat{\sigma}_{ij}^{\rm tr'}}{d\xi_p}$
Eq.~(\ref{eq:smallptjoint}). 
Note that in the
$\pt\to0$ limit the cross section starts at $\Ord(\alpha_s^0)$ (in the
effective field theory limit, corresponding to $\Ord(\alpha_s^2)$ in
full QCD with heavy quark masses), while for finite $\pt$ the
transverse momentum distribution starts at $\Ord(\alpha_s)$, hence
there is a mismatch in counting fixed or resummed order: e.g. a
N$^k$LO contribution to the transverse momentum spectrum is a
N$^{k+1}$LO contribution to the $\pt\to0$ (or inclusive)
cross section. However, because of the way our combined resummed
expression Eq.~(\ref{eq:jointfinal}) is constructed there is no harm
in including either of the two contributions to it at an extra
perturbative order. Note that  two different definitions of logarithmic
order exist in the literature, according to whether one counts orders
in the exponent, or for the expanded cross section; we will follow the
nomenclature given e.g. in Table~1 of
Ref.~\cite{Bonvini:2014qga}.

We will present results up to NNLL accuracy for
transverse momentum resummation and to NNLL* for threshold resummation
at finite $\pt$, which, upon integration
over $\xi_p$ reproduce  NNLL*
threshold resummation for the inclusive cross section. 
Explicit values of the
coefficients needed in order to achieve this accuracy are collected in
Appendices~\ref{subapp:largeN}-\ref{subapp:smallpt}, while in
Appendix~\ref{subapp:inclN} we collect the coefficients which are
needed in order to check that inclusive threshold resummation is
reproduced.  For transverse momentum resummation, they are obtained
matching our generalized resummed result
$\frac{d\hat{\sigma}_{ij}^{\rm tr'}}{d\xi_p}$ Eq.~(\ref{eq:smallptjoint})
to the NNLL result of
Refs.~\cite{Bozzi:2005wk,Catani:2011kr}. For threshold resummation they are
obtained by extending up to  NNLL* the NLL expression of
Ref.~\cite{deFlorian:2005fzc}.

As already mentioned, Eq.~\eqref{eq:jointfinal} reproduces by
construction threshold 
resummation in the large-$N$ limit at fixed
$\xi_p$, and transverse 
momentum resummation Eq.~(\ref{eq:smallptjoint}) in the $\pt\to 0$ limit
at fixed $N$ and it reproduces the threshold limit of the total
cross section upon integration over $\pt$ if the modified transverse
momentum resummation $\frac{d\hat{\sigma}_{ij}^{\rm tr'}}{d\xi_p}$ does. 
It thus remains to work out the  explicit 
expression for the contribution $\frac{d\hat{\sigma}_{ij}^{\rm
    tr'}}{d\xi_p}$ Eq.~(\ref{eq:smallptjoint}), and check that
computing it with the values of the
coefficients listed in Appendix~\ref{subapp:smallpt}, such that it
gives standard transverse momentum resummation  
Eq.~(\ref{eq:sudpt}) in the $\pt\to0$ limit, it indeed leads to 
threshold resummation
of the total cross section upon integration over $\pt$. As a byproduct
of this check, we will end up with an explicit closed-form expression
for the combined resummed result.

We start from the resummed expression
Eq.~\eqref{eq:smallptjoint}. As discussed
above, the exponent of 
this expression ought to be evaluated at leading order in an
expansion in powers of $\frac{1}{b}$ at  fixed $\frac{N}{b}$. As a
consequence of our combined resummation formalism, this expression, 
as a function of $N$ and $b$, reduces to standard
transverse momentum resummation if expanded to leading order in powers
of $b$ at fixed $N$, while for
$b=0$ it reproduces the total  cross section in the threshold limit $N
\to \infty$. 

Expanding
$A^{\pt}$ and  $\mathcal{B}$ in powers of $\as$ and expressing the result
in terms of  $\as\(M^2\)$ all integrals in the exponent are of the form
\begin{align}
\label{eq:G1}
&G_{k,1}\(N,b\)=\int_0^{\infty} d\xi \(\sqrt{1+\xi}-\sqrt{\xi}\)^{2N} J_0\(b M\sqrt{\xi}\)\int_0^1 dz\,z^{N-1}\notag\\
&\qquad \Bigg(\plus{\frac{\ln^k \xi}{\xi}}^{\pt} \plus{\frac{1}{\sqrt{\(1-z\)\(1-\(\sqrt{1+\xi}-\sqrt{\xi}\)^4 z\)}}}^z\notag \\
&\qquad+\delta\(1-z\)\frac{1}{2\(\sqrt{1+\xi}-\sqrt{\xi}\)^2}\notag\\
&\qquad\qquad\qquad\quad\( \frac{\ln\(1+\xi\)\ln^k\xi}{\xi}-\plus{\frac{\ln^{k+1}\xi}{\xi}}^{\pt}\)\Bigg)\\
\label{eq:G2}
&G_{k,2}\(N,b\)=\int_0^{\infty} d\xi\, \(\sqrt{1+\xi}-\sqrt{\xi}\)^{2N} J_0\(b M \sqrt{\xi}\) \plus{\frac{\ln^k\xi}{\xi}}^{\pt},
\end{align}
where the integrals $G_{k,1}$ and $G_{k,2}$ appear in the terms
proportional to $A^{\pt}$ and  $\mathcal{B}$ respectively. We need to
evaluate these integrals in the limit $b\to \infty$ at fixed $\frac{N}{b}$.

This can be done by defining 
two generating functions, $\mathcal{G}_1\(N, b,\epsilon\)$ and $\mathcal{G}_2\(N,b,\epsilon\)$ such that
\begin{align}\label{eq:Gdiff}
G_{k,1}\(N,b\)&=
\left.\frac{d^k}{d\epsilon^k} \mathcal{G}_1\(N,b,\epsilon\)\right|_{\epsilon=0}
\nonumber \\
G_{k,2}\(N,b\)&=
\left.\frac{d^k}{d\epsilon^k} \mathcal{G}_2\(N,b,\epsilon\)\right|_{\epsilon=0}.
\end{align}
These admit  the  integral representation
\begin{align}
\label{eq:Gcal1}
\mathcal{G}_1\(N, b, \epsilon\)&=\int_0^1 dz\,z^{N-1}\int_0^{\frac{\(1-z\)^2}{4z}} d\xi\,J_0\(b M \sqrt{\xi}\) \frac{\xi^{-1+\epsilon}}{\sqrt{\(1-z\)^2-4 z \xi}}
\notag\\
&-\frac{1}{2\epsilon^2}-\frac{1}{\epsilon}\int_0^1 dz\,z^{N -1} \plus{\frac{1}{1-z}}\\
\label{eq:Gcal2}
\mathcal{G}_2\(N, \epsilon\)&=\int_0^\infty d\xi\,\left[\(\sqrt{1+\xi}-\sqrt{\xi}\)^{2N}\, J_0\(b M \sqrt{\xi}\)-1\right] \xi^{-1+\epsilon}.
\end{align}
Expanding the Bessel function in powers of its argument 
\beq
J_0\(b M \sqrt{\xi}\)=\sum_{p=0}^\infty 
\frac{(-1)^p}{\Gamma^2\(p+1\)}\(\frac{b^2 M^2}{4}\)^p \xi^p
\eeq
and integrating term by term we get
\begin{align}
\label{eq:G1cal2}
\mathcal{G}_1\(N, b, \epsilon\)&=
\sum_{p=0}^\infty \left[\frac{(-1)^p}{\Gamma^2\(p+1\)}\(\frac{b^2 M^2}{4}\)^p 
\frac{\Gamma\(N - p - \epsilon\) \Gamma^2\(p+\epsilon\)}
{2\Gamma\(N+p +\epsilon\)}\right]
\notag\\
&-\frac{1}{2\epsilon^2}+\frac{1}{\epsilon}\(\psi\(N\)+\gammae\)
\\
\label{eq:G2cal2}
\mathcal{G}_2\(N, b, \epsilon\)&=
\sum_{p=0}^\infty \left[\frac{(-1)^p}{\Gamma^2\(p+1\)}\(\frac{b^2 M^2}{4}\)^p 
\frac{N \Gamma\(N - p-\epsilon\)\Gamma\(2\(p+\epsilon\)\)}
{2^{2p+2\epsilon-1} \Gamma\(N+1 + p +\epsilon\)}\right]\notag\\
&-\frac{1}{\epsilon}.
\end{align} 
We can now take the large-$b$ limit at fixed $\frac{N}{b}$. 
Because the $\Gamma$ functions do not depend on $b$,
this limit can be taken using the asymptotic expansion
\beq
\label{eq:asimgamma}
\frac{\Gamma\(N-p-\epsilon\)}{\Gamma\(N+p+\epsilon\)}
=\(\frac{1}{N^2}\)^{p+\epsilon}\(1+ \Ord\(\frac{1}{N}\)\).
\eeq
By inserting Eq.~\eqref{eq:asimgamma} into
Eqs.~(\ref{eq:G1cal2},\ref{eq:G2cal2}) and performing the sum on $p$ we obtain
\begin{align}
\label{eq:G1calfinal}
\mathcal{G}_1\(N, b,\epsilon\)&=\frac{1}{2} \(\frac{1}{N^2}\)^{\epsilon} 
\Gamma^2\(\epsilon\) \,_2F_1\(\epsilon,\epsilon,1,-\frac{b^2M^2}{4N^2}\)-\frac{1}{2\epsilon^2}
+\frac{1}{\epsilon}\(\ln N+\gammae\)+\Ord\(\frac{1}{b}\)\\
\label{eq:G2calfinal}
\mathcal{G}_2\(N, b, \epsilon\)&=2^{1-2\epsilon}\(\frac{1}{N^2}\)^{\epsilon} \Gamma\(2\epsilon\) \,_2F_1\(\epsilon,\frac{1}{2}+\epsilon,1,-\frac{b^2M^2}{4N^2}\)-\frac{1}{\epsilon}+\Ord\(\frac{1}{b}\).
\end{align}
These provide us with the desired expressions of the generating
functions at leading order in the $b\to\infty$ limit, for fixed $\frac{N}{b}$.

The derivatives of the generating functions
Eqs.~(\ref{eq:G1calfinal}-\ref{eq:G2calfinal}) could be performed
using recent results~\cite{Huber:2005yg,Huber:2007dx} for the
expansion of
hypergeometric function in powers of $\epsilon$. However, very compact
closed-form expressions can be obtained by replacing the generating functions
Eqs.~(\ref{eq:G1calfinal}-\ref{eq:G2calfinal}) with suitable expressions
which only differ by them by subleading terms.
Indeed, because powers of $\ln
\xi$ are obtained by differentiation with respect to $\epsilon$
according to Eq.~(\ref{eq:Gdiff}),
an expression of the
generating functions which reproduces transverse momentum resummation
up to N$^k$LL accuracy can  be obtained by expanding the
hypergeometric functions in powers 
of $\epsilon$, and at each order in $\epsilon$ evaluating its large
$b$ limit and  retaining the $k+1$
highest powers of $\ln b$. Furthermore, because of the prefactor of
$N^{-2\epsilon}$ in Eqs.~(\ref{eq:G1calfinal}-\ref{eq:G2calfinal}) 
 an expression of the
generating functions which reproduces inclusive threshold momentum resummation
up to N$^j$LL accuracy can be obtained by letting $b=0$ and then 
expanding the
hypergeometric functions in powers 
of $\epsilon$ and retaining the first $j$ orders of the
expansion. Hence, any function which reproduces these two  behaviors
of the original generating functions will lead to the same resummed
results to the desired accuracy: up to NNLO this requires  $k=j=2$.

For $\mathcal{G}_1$, we do this by noting that the hypergeometric function
${}_2F_1$ has the asymptotic expansion for large $z$
\beq\label{eq:hypasympt1}
_2F_1\(\epsilon,\epsilon,1,-z\)=\frac{z^{-\epsilon}}{\Gamma\(\epsilon\)\Gamma\(1-\epsilon\)}\(\ln z- \psi\(1-\epsilon\)-\psi\(\epsilon\)-2\gammae\)+\Ord\(\frac{1}{z}\),
\eeq
and the Taylor expansion
\beq\label{eq:hyptaylor}
{}_2F_1\(\epsilon,\epsilon,1,-z\)=1+\epsilon^2 \Li_2\(-z\)+\Ord\(\epsilon^3\).
\eeq
We can easily combine these two behaviours  by first, letting
$z\to1+z$ on the right-hand side of Eq.~(\ref{eq:hypasympt1}): this
leads to an expression which coincides with Eq.~(\ref{eq:hypasympt1})
as $z\to\infty$ up to $\Ord\left(\frac{1}{z}\right)$ corrections, but
is regular as $z\to0$. Next, we expand the result in powers of
$\epsilon$ and we match to the expansion
Eq.~(\ref{eq:hyptaylor}). Namely, we note that
\begin{equation}\label{eq:secexp}
\frac{(1+z)^{-\epsilon}}{\Gamma\(\epsilon\)\Gamma\(1-\epsilon\)}\(\ln(1+ z)- \psi\(1-\epsilon\)-\psi\(\epsilon\)-2\gammae\)=1-\epsilon^2\left[\frac{1}{2}\ln^2(1+z)+\zeta_2\right]+\Ord\(\epsilon^3\).
\end{equation}
But 
\beq\label{eq:diff}
\Li_2\(-z\)+\frac{1}{2}\ln^2\(1+z\)+\zeta_2=\Li_2\(\frac{1}{1+z}\)-\(\ln\(1+z\)-\ln\(z\)\)\ln\(1+z\).
\eeq
Hence it is enough to add the left-hand side of
Eq.~(\ref{eq:diff}) to the right-hand side of
Eq.~(\ref{eq:hypasympt1}) after having performed in it the $z\to1+z$ 
shift, to get an
interpolation of the hypergeometric function which, if substituted in
Eq.~(\ref{eq:G1calfinal}), leads to the same result to up to
subleading power corrections in the small $\pt$ limit and up to N$^3$LL
corrections in the threshold limit at the integrated level. This
can be increased to N$^j$LL by including the expansion in powers of
$\epsilon$ in Eqs.~(\ref{eq:hyptaylor},\ref{eq:secexp}) up to $j-1$--th
order. 

The second
term on the right-hand side of Eq.~(\ref{eq:diff}) can be dropped, as
it is $O(z)$ as $z\to0$ and $O\(\frac{1}{z}\)$ as $z\to\infty$,
and so we end up with
the result
\begin{align}\label{eq:finasymptg1}
_2F_1\(\epsilon,\epsilon,1,-z\)&=\frac{\(1+z\)^{-\epsilon}}{\Gamma\(1-\epsilon\)\Gamma\(\epsilon\)}\(\ln\(1+z\)-2\gammae-\psi\(1-\epsilon\)-\psi\(\epsilon\)\)\notag\\
&+\epsilon^2 \Li_2\(\frac{1}{1+z}\)+\Ord\(\rm NNNLL\),
\end{align}
where the order of the correction means that using
Eq.~(\ref{eq:finasymptg1}) in the expression Eq.~(\ref{eq:G1calfinal}) of the
generating
 function $\mathcal{G}_1$ leads to resummed  expression which preserve the
original accuracy in the small $\pt$ limit, and which are NNLL*
accurate in the threshold limit upon integration over $\pt$.

For $\mathcal{G}_2$ we use the expansion
\begin{equation}\label{eq:hypasympt2}
_2F_1\(\epsilon,\frac{1}{2}+\epsilon,1,-z\)=\frac{\sqrt{\pi}2^{-2\epsilon}
    z^{-\epsilon}}{\Gamma\(\frac{1}{2}+\epsilon\)\Gamma\(1-\epsilon\)}+\Ord\(\frac{1}{z}\).
\end{equation}
We note furthermore that $\mathcal{G}_2$ generates the integrals which
enter in the terms proportional to $\mathcal{B}$ in the resummed
expression Eq.~(\ref{eq:smallptjoint}). These start at NLL, hence, up
to NNLL accuracy, it
is sufficient to perform the expansion in powers of $\epsilon$ up
to first order, rather than second order as in
Eq.~(\ref{eq:hyptaylor}). Furthermore, we note that the
$\Ord(\epsilon)$ term in this expansion only receives a contribution
from the leading-order contribution to  $\mathcal{B}$, which vanishes
in the threshold limit $N\to\infty$, see Eq.~(\ref{eq:calB1}). It
follows that is enough to reproduce the expansion
\beq\label{eq:epsg2}
\,_2F_1\(\epsilon,\epsilon,1,-z\)=1+\Ord(\epsilon).
\eeq
This  is automatically the case is we simply perform
the shift $z\to 1+z$ on the right-hand side of Eq.~(\ref{eq:hypasympt1}).
We thus end up with
the result
\beq\label{eq:finasymptg2}
_2F_1\(\epsilon,\frac{1}{2}+\epsilon,1,-z\)=\frac{\sqrt{\pi}2^{-2\epsilon}
    \(1+z\)^{-\epsilon}}{\Gamma\(\frac{1}{2}+\epsilon\)\Gamma\(1-\epsilon\)}
+\Ord\(\rm NNNLL\),
\eeq
where again the order of the correction means that using this result
in 
Eq.~(\ref{eq:G2calfinal}) 
leads to resummed  expression which preserve the
original accuracy in the small $\pt$ limit, and which are NNLL*
accurate in the threshold limit upon integration over $\pt$.

Inserting the expanded 
expressions Eqs.~(\ref{eq:finasymptg1},\ref{eq:finasymptg2}) into Eq.~\eqref{eq:G1},~\eqref{eq:G2} and performing the derivatives, we obtain, up to NNLL
\begin{align}
G_{k,1}\(N,b\)&=\frac{\(-1\)^k}{2}\Bigg[-\frac{1}{k+2}\ln^{k+2}{\chi} + \frac{\ln \bar{N}^2}{k+1} \ln^{k+1} {\chi}+ \ln^k \bar{N}^2\;\Li_2\(\frac{\bar{N}^2}{{\chi}}\)\notag\\
&+\Ord\(\ln^{j} \bar{N}^2\,\ln^{k-1-j} {\chi}\)\Bigg]\\
G_{k,2}\(N,b\)&=-\frac{(-1)^k}{k+1}\ln^{k+1} {\chi} + 
\Ord\(\ln^{k-1} {\chi}\)
\end{align}
with 
\begin{align}\label{eq:nbardef}
\bar{N}&=N\,e^{\gammae} \\\label{eq:chidef} 
{\chi} &= \bar{N}^2+ \frac{b^2 M^2}{b_0^2}\\\label{eq:b0def}
b_0 &=2e^{-\gammae}.
\end{align}

Using this result, we can cast Eq.~\eqref{eq:smallptjoint} in the
familiar form
\begin{align}
\label{eq:smallptjointfinal}
&\frac{d\hat{\sigma}_{ij}^{\rm tr'}}{d\xi_p}\(N,\xi_p,\as\(M^2\),M^2\)=\sigma_0 \int_0^{\infty} db\, \frac{b}{2}\, J_0\(b M \sqrt{\xi_p}\) \(\sqrt{1+\xi_p}-\sqrt{\xi_p}\)^{-2N}\notag\\
&\bar{\mathcal{H}}_{ij}\(N,\as\(M^2\)\) \exp\left[\ln {\chi}\,g_1\(\lN,\lchi\)+g_2\(\lN,\lchi\)+\alpha\,g_3\(\lN,\lchi\)\right]\notag\\
&+\Ord\(\rm NNNLL\).
\end{align}
This is our main result to be used for applications.
The functions $g_1, g_2, g_3$ resum the LL, NLL, NNLL
contributions respectively,  and  depend on the two large resummation
logs
\begin{align}\label{eq:largelogN}
 \lN&=\as\(M^2\) \beta_0 \ln \bar{N}^2\\ \label{eq:largelogb}
\lchi&=\as\(M^2\)\beta_0\ln {\chi}.
\end{align}
Note that the function ${\chi}$ interpolates between $b^2$ at large $b$,
and $N^2$ when $b=0$:
\begin{align}\label{eq:blimsa}
{\chi}&=\frac{b^2  M^2}{b_0^2}
\[1+\Ord\(\frac{1}{b^2}\)\]\\
\label{eq:blimsb}
{\chi}&=\bar{N}^2\left[1+\Ord\(\frac{1}{N^2}\)\],
\end{align}
 hence, in the former limit,  the form of
Eq.~(\ref{eq:largelogN}) naturally matches the standard expression of
transverse momentum resummation of Ref.~\cite{Bozzi:2005wk}.

Evaluating the functions  $g_i$ up to NNLL explicitly we get
\small
\label{eq:gi}
\begin{align}
\label{eq:g1}
g_1\(\lchi,\lN\)&=\frac{A^{\pt, (1)}_g}{\beta_0}\(\frac{\lchi+\ln\(1-\lchi\)}{\lchi}\)-\frac{A^{\pt, (1)}_g}{\beta_0}\ln\(1-\lchi\)\frac{\lN}{\lchi}\\
\label{eq:g2}
g_2\(\lchi,\lN\)&=\frac{A^{\pt, (1)}_g \beta_1}{\beta_0^3}\left[\frac{\lchi+\ln\(1-\lchi\)}{1-\lchi}+\frac{1}{2}\ln^2\(1-\lchi\)\right]\notag\\
&-\frac{A^{\pt, (2)}_g}{\beta_0^2}\left[\ln\(1-\lchi\)+\frac{\lchi}{1-\lchi}\right] +\frac{\mathcal{B}^{(1)}_g\(N\)}{\beta_0}\ln\(1-\lchi\)\notag\\
&-\frac{A^{\pt, (1)}_g \beta_1 \lN}{\beta_0^3}\(\frac{\lchi+\ln\(1-\lchi\)}{1-\lchi}\)+\frac{A^{\pt, (2)}_g \lN}{\beta_0^2}\frac{\lchi}{1-\lchi}\\
\label{eq:g3}
g_3\(\lchi,\lN\)&=\frac{A^{\pt, (1)}_g \beta_1^2}{2\beta_0^4}\left[\frac{\lchi+\ln\(1-\lchi\)}{\(1-\lchi\)^2}\(\lchi+\(1-2\lchi\)\ln\(1-\lchi\)\)\right]\notag\\
&+\frac{A^{\pt, (1)}_g \beta_2}{\beta_0^3}\left[\frac{\(2-3\lchi\)\lchi}{2\(1-\lchi\)^2}+\ln\(1-\lchi\)\right]\notag\\
&-\frac{A^{\pt, (2)}_g \beta_1}{\beta_0^3}\left[\frac{\(2-3\lchi\)\lchi}{2\(1-\lchi\)^2}+\frac{\(1-2\lchi\)\ln\(1-\lchi\)}{\(1-\lchi\)^2}\right]\notag\\
&+\frac{\mathcal{B}^{(1)}_g\(N\) \beta_1}{\beta_0}\frac{\lchi+\ln\(1-\lchi\)}{1-\lchi}-\frac{A^{\pt, (3)}_g \lchi^2}{2\beta_0^2 \(1-\lchi\)^2}-\frac{\mathcal{B}_g^{(2)}\(N\)}{\beta_0} \frac{\lchi}{1-\lchi}\notag\\
&+A^{\pt, (1)}_g \frac{\lN}{1-\lN} \Li_2\(\frac{\bar{N}^2}{{\chi}}\)-\frac{A^{\pt, (1)}_g \beta_1^2\,\lN}{2\beta_0^4}\frac{\lchi^2-\ln^2\(1-\lchi\)}{\(1-\lchi\)^2}\notag\\
&-\frac{A^{\pt, (2)}_g \beta_1\,\lN}{2\beta_0^3}\frac{\lchi\(2-\lchi\)+2\ln\(1-\lchi\)}{\(1-\lchi\)^2}\notag\\
&+\frac{A^{\pt, (3)}_g}{2\beta_0^2} \frac{\lN \lchi\(2-\lchi\)}{\(1-\lchi\)^2}+\frac{A^{\pt, (1)}_g \beta_2}{2\beta_0^3}\frac{\lN \lchi^2}{\(1-\lchi\)^2}
\end{align}
\normalsize
while the hard function is
\beq
\label{eq:Hmod}
\bar{\mathcal{H}}_{ij}\(N,\as\(M^2\)\)=\mathcal{H}_{ij}\(N,\as\(M^2\)\) +\delta_{ij = g}\,A_g^{\pt, (1)} \, \Li_2\(\frac{\bar{N}^2}{{\chi}}\).
\eeq
It is then immediate to see that, replacing ${\chi}$
  with its limiting large $b$ form Eq.~(\ref{eq:blimsa}) and using the
  explicit expressions of all the coefficients given in
  Appendix~\ref{subapp:smallpt} we recover the expression for
  transverse momentum resummation of Ref.~\cite{Bozzi:2005wk} (see
  specifically Eqs.~(22-24) of that reference). 

We can now proceed to the nontrivial consistency check of our
procedure, namely, that setting $b=0$ we recover threshold resummation
at the inclusive level up to NNLL* accuracy. Up to NLL accuracy,
inclusive threshold resummation is entirely determined by the cusp
anomalous dimension  $A_g^{\rm th}$ [explicitly given in
Eqs.~(\ref{eq:cusp}-\ref{eq:cusp3})], namely,
the coefficient of the
most singular contribution to the anomalous dimension as $N\to\infty$,
which in the $\overline{\rm MS}$ scheme is proportional to $\ln
N$~\cite{Albino:2000cp}. It follows that threshold resummation is
reproduced automatically up to this order if the coefficients of the
expansion of the
function $A_g^{\rm \pt}$ Eq.~(\ref{eq:sudpt}) coincide with those of
the cusp anomalous dimension:
\begin{align}\label{eq:match1}
A_g^{\rm \pt, (1)}&=A_g^{\rm th, (1)}\\\label{eq:match2}
A_g^{\rm \pt, (2)}&=A_g^{\rm th, (2)},
\end{align}
which is of course the case. Note that this fact is nontrivial: indeed
 $A_g^{\rm \pt}$ is implicitly defined by Eq.~(\ref{eq:sudpt}), which
determines order 
  by order the way resummation coefficients are assigned to the
  functions $A_g^{\rm \pt}$ and $B_g^{\rm \pt}$.  Our derivation shows
  that this effectively amounts to defining $A_g^{\rm \pt}$ as the
  function which includes in the Sudakov exponent terms
  which are enhanced by $\ln N$, i.e., that controls the resummation of
  terms proportional to $\ln N\ln b$, which, beyond the leading log
  level, is not obviously the same as the cusp anomalous dimension.

And indeed, starting at the NNLL it is not. At this order and beyond, threshold
resummation at the inclusive level also receives a contribution from
large-angle gluon emission~\cite{Catani:2001ic}, described by an
additional function   $D_g^{\rm \pt}$ (see
Eqs.~(\ref{eq:thrres1}-\ref{eq:largang2}). Up  to NNLL* accuracy we then
recover inclusive threshold resummation
setting $b=0$ in Eq.~(\ref{eq:smallptjointfinal}) only if the
following relations are satisfied:
\begin{align}
\label{eq:collanomaly} 
A_g^{\rm \pt, (3)}+\beta_0 D_g^{\rm \pt, (2)}&=A_g^{\rm th, (3)}\\
\label{eq:match4}
D_g^{\rm \pt, (2)}+2 \tilde{B}_g^{\rm \pt, (2)}+2A_g^{\rm \pt, (1)} \zeta_2 \beta_0&=D_g^{\rm th, (2)}\\\label{eq:match5}
H_{gg}^{\rm \pt, (1)}+A_g^{\rm \pt, (1)} \zeta_2&= H_{gg}^{\rm th, (1)},
\end{align}
where the  coefficients on the 
 right-hand side of Eqs.~(\ref{eq:collanomaly}-\ref{eq:match5}),
 which determine NNLL threshold resummation at the inclusive level, are
explicitly listed in Appendix~\ref{subapp:inclN}. It is
straightforward to check that Eqs.~(\ref{eq:match1}-\ref{eq:match5})
are indeed satisfied. The fact that at NNLL and beyond $A_g^{\rm \pt}$
does not coincide with the cusp anomalous dimension was first shown
using SCET arguments in Ref.~\cite{Becher:2010tm,Becher:2012yn}, where
it was derived from the breaking of a symmetry of the classical SCET
Lagrangian called
``collinear anomaly''. We
now see that this simply means that the coefficient of $\ln N$ in the
anomalous dimension, and the coefficient  of $\ln N\ln b$ in
transverse momentum
resummation, do not coincide, because the latter receives contribution from
interference between  soft virtual corrections, controlled by
$D^{\pt}_g$, and collinear emission.  

Note that the accuracy of our results is NNLL*, rather than  NNLL,
because  $H_{gg}^{\rm \pt, (2)}$  differs from 
$H_{gg}^{\rm th, (2)}$ by terms proportional to $g_4$, which we do not
fully include.
All results presented here apply to Higgs production in gluon fusion;
however, results for Drell-Yan production have the same structure and
are obtained by simply replacing the expressions of the functions
$A^{\pt}_g$ and
$\mathcal{B}_g\(N\)$ with their quark counterparts $A^{\pt}_q$ and
$\mathcal{B}_q\(N\)$.

\section{Conclusions}
\label{sec:conclusion}
In this paper we have constructed an expression for the transverse
momentum distribution of a colorless object in perturbative QCD which
reduces to transverse momentum resummation in the small $\pt$ limit at
fixed $x$ and to threshold resummation in the $x\to1$ limit at fixed
$\pt$, and which, furthermore, gives threshold resummation at the
inclusive level when integrated over transverse momentum. Our combined
resummed expression is
the matched formula Eq.~(\ref{eq:jointfinal}) with
Eqs.~(\ref{eq:tlargen}-\ref{eq:tsmallxi}). Its main original
ingredient is the modified transverse momentum resummation formula,
given as a master integral formula in Eq.~(\ref{eq:smallptjoint}) and
as a compact closed-form expression
in Eq.~(\ref{eq:smallptjointfinal}). In fact,
Eq.~(\ref{eq:smallptjointfinal})
is our main new result:
it provides a modified expression for transverse momentum resummation
which automatically reproduces  threshold resummation upon integration over $p_T$.

The interest in this construction is threefold. First, we provide
resummed expressions which can be used for phenomenology at a 
differential level, allowing for an improvement of the transverse
momentum distribution through threshold resummation in a way that
holds for all values of $\pt$ and which matches onto  inclusive
results which have been similarly improved.  Second, our results, when
expanded out to finite order in $\alpha_s$ provide powerful
constraints on higher-order perturbative corrections, which may be
used as consistency check of full calculations, and as a means of
constructing approximate results for yet unknown higher order
terms. Finally, we elucidate the relation between the collinear and
soft logs which drive the transverse momentum distribution and the
total cross section in the soft limit.

The virtues of our final result can be perhaps best understood by
comparing it to other related results which have been previously derived.
In Ref.~\cite{Kulesza:2003wn} a joint resummation for Higgs
production  up to NLL was derived, by  studying 
singular eikonal emission within the web formalism. A resummed result 
was obtained in terms of an interpolating function 
$\bar{\chi}=\frac{b M}{b_0}+\bar{N}$ which can be compared to
our resummation log $\lchi$ Eq.~(\ref{eq:largelogb}) with
${\chi}$ Eq.~(\ref{eq:chidef}). This result was recently
extended up to NNLL in Ref.~\cite{Marzani:2016smx} by means of a
suitable Ansatz. 
We have
checked that our result reproduces that of these references in the sense
that our $g_1$ and $g_2$ resummation functions
Eqs.~(\ref{eq:g1},\ref{eq:g2})
coincide with the corresponding expressions of
Ref.~\cite{Kulesza:2003wn} and  $g_3$  Ref.~\cite{Marzani:2016smx} 
once differences in notation are accounted
for. However, the logs of the interpolating function
$\bar{\chi}$ of Ref.~\cite{Kulesza:2003wn} produce 
terms which are subleading in the small $\pt$ limit but induce in
the result of Ref.~\cite{Kulesza:2003wn}   
unphysical logs of  $\frac{N}{b}$ which are not present at any finite order.
In order to remove them, a phenomenological modification of $\chi$ was 
proposed in
Ref.~\cite{Laenen:2000ij} and used in Ref.~\cite{Kulesza:2003wn},
$\bar{\chi}_{\rm phen}\(\eta\)=\frac{b M}{b_0}+\frac{\bar{N}}{1+\eta
  \frac{b M}{2 N}}$, dependent on a free parameter $\eta$  and such
that if  $\eta\not=0$ the spurious large $N$ behaviour is removed up
to order $b^{-1}$. Our result avoids these ad-hoc manipulations.
Finally, in Ref.~\cite{Lustermans:2016nvk}, a NNLL joint resummation
in SCET was performed. This resummation is directly
performed in $x$ and $\pt$ space, and it does not appear to reproduce
threshold resummation at the inclusive level upon integration over
transverse momentum.

Besides phenomenological studies, future directions of progress include 
the possibility of merging  the
result of this paper with the recent high energy resummation for
transverse momentum distributions performed at 
fixed-$\pt$ in Ref.~\cite{Forte:2015gve,Caola:2016upw}, and already matched to transverse
momentum resummation in  Ref.~\cite{Marzani:2015oyb}, with the
eventual goal of deriving resummation of the fully exclusive
cross section in all kinematic limits.
  
{\bf Acknowledgements:} We are grateful to Fabrizio Caola for a critical reading
of the manuscript and several interesting observations. We also thank
Giancarlo Ferrera and
Simone Marzani for useful discussions.

\appendix
\section{Explicit expressions}
\label{app:explicit}
\subsection{Threshold resummation at fixed $\pt$}
\label{subapp:largeN}

We give here explicit expressions of the
coefficients which determine threshold resummation for Higgs production
in gluon fusion in the pointlike limit up to
NNLL* accuracy when used in the expression Eq.~(\ref{eq:largeNjoint})
of $\frac{d\hat{\sigma}_{ij}^{\rm th}}{d\xi_p}$.
The cusp anomalous dimensions $A^{\rm th}_g\(\as\)$ and $A^{\rm th}_q\(\as\)$ i.e. 
the contribution to the $P_{gg}$ and $P_{qq}$ splitting
function respectively 
which  are proportional to a plus distribution are given by
(see e.g. Ref.~\cite{Moch:2005ba}):
\begin{align}\label{eq:cusp}
&A^{\rm th}_c\(\as\)= A^{\rm th, (1)}_c \(\frac{\as}{\pi}\)+ A^{\rm th, (2)}_c \(\frac{\as}{\pi}\)^2 + A^{\rm th, (3)}_c \(\frac{\as}{\pi}\)^3 + \Ord\(\as^4\)\notag\\
&\quad\notag\\
&A^{\rm th, (1)}_c=C_c,\\
&A^{\rm th, (2)}_c=\frac{C_c}{2}\Bigg[\Ca\(\frac{67}{18}-\zeta_2\)-\frac{5}{9}\Nf\Bigg],\\\label{eq:cusp3}
&A^{\rm th, (3)}_c=C_c\Bigg[\(\frac{245}{96}-\frac{67}{36}\zeta_2+\frac{11}{8}\zeta_4+\frac{11}{24}\zeta_3\)\Ca^2+\(-\frac{209}{432}+\frac{5}{18}\zeta_2-\frac{7}{12}\zeta_3\)\Ca\Nf\notag\\
&\qquad\qquad\qquad+\(-\frac{55}{96}+\frac{1}{2}\zeta_3\)\Cf \Nf-\frac{1}{108}\Nf^2\Bigg],
\end{align}
 with $C_c=\Ca$ if $c=g$ is a gluon and $C_c=\Cf$ if
 $c=q$. Furthermore 
\begin{align}
&B^{\rm th}_c\(\as\)=B^{\rm th, (1)}_c \(\frac{\as}{\pi}\)+ B^{\rm th, (2)}_c \(\frac{\as}{\pi}\)^2+ \Ord\(\as^3\)\notag\\
&\quad\notag\\
&B^{\rm th, (1)}_q=-\frac{3}{4}\Cf,\\
&B^{\rm th, (2)}_q=\frac{1}{16}\Bigg[\Cf^2\(-\frac{3}{2}+12\zeta_2-24\zeta_3\)+\Cf\Ca\(-\frac{3155}{54}+\frac{44}{3}\zeta_2+40\zeta_3\)\notag\\
&\qquad\qquad\qquad+\Cf\Nf\(\frac{247}{27}-\frac{8}{3}\zeta_2\)\Bigg],\\
\,\notag\\
&B^{\rm th, (1)}_g=-\beta_0=-\frac{11}{12}\Ca+\frac{1}{6}\Nf,\\
&B^{\rm th, (2)}_g=\frac{1}{16}\Bigg[\Ca^2\(-\frac{611}{9}+\frac{88}{3}\zeta_2+16\zeta_3\)+\Ca\Nf\(\frac{428}{27}-\frac{16}{3}\zeta_2\)+2\Cf\Nf-\frac{20}{27}\Nf^2\Bigg].
\end{align}
Note that both quark and gluon channel expressions are necessary for
NNLL* accuracy.

We now turn to the LO coefficient functions$C_0\(N,\xi_p\)$ in
Eq.~\eqref{eq:largeNjoint}: after factoring the leading-order total
cross section
\beq
\label{eq:sigma0}
\sigma_0=\frac{\as^2 \sqrt{2} \Gf}{576\pi}.
\eeq
they are given by
\footnotesize
\begin{align}
&\frac{d\hat{\sigma}^{\rm LO}_{gg \to g H}}{d\xi_p}\(N,\xi_p\)=\frac{2\as\Ca}{\pi}\frac{1}{\xi_p}\frac{\Gamma\(\frac{1}{2}\)\Gamma\(N\)}{\Gamma\(N+\frac{1}{2}\)}\Bigg( \,_2F_1\(\frac{1}{2},N,N+\frac{1}{2},\(\sqrt{1+\xi_p}-\sqrt{\xi_p}\)^4\)\notag\\
&-2\frac{1+\xi_p}{\(\sqrt{1+\xi_p}+\sqrt{\xi_p}\)^2}\frac{N}{N+\frac{1}{2}}\,_2F_1\(\frac{1}{2},N+1,N+\frac{3}{2},\(\sqrt{1+\xi_p}-\sqrt{\xi_p}\)^4\)\notag\\
&+\frac{\(1+\xi_p\)\(3+\xi_p\)}{\(\sqrt{1+\xi_p}+\sqrt{\xi_p}\)^4}\frac{N\(N+1\)}{\(N+\frac{1}{2}\)\(N+\frac{3}{2}\)} \,_2F_1\(\frac{1}{2},N+2,N+\frac{5}{2},\(\sqrt{1+\xi_p}-\sqrt{\xi_p}\)^4\)\notag\\
&-2\frac{1+\xi_p}{\(\sqrt{1+\xi_p}+\sqrt{\xi_p}\)^6} \frac{N\(N+1\)\(N+2\)}{\(N+\frac{1}{2}\)\(N+\frac{3}{2}\)\(N+\frac{5}{2}\)} \,_2F_1\(\frac{1}{2},N+3,N+\frac{7}{2},\(\sqrt{1+\xi_p}-\sqrt{\xi_p}\)^4\)\notag\\
&+\frac{1}{\(\sqrt{1+\xi_p}+\sqrt{\xi_p}\)^8}\frac{N\(N+1\)\(N+2\)\(N+3\)}{\(N+\frac{1}{2}\)\(N+\frac{3}{2}\)\(N+\frac{5}{2}\)\(N+\frac{7}{2}\)}\notag\\
& \,_2F_1\(\frac{1}{2},N+4,N+\frac{9}{2},\(\sqrt{1+\xi_p}-\sqrt{\xi_p}\)^4\)\Bigg)\\
&\,\notag\\
&\frac{d\hat{\sigma}^{\rm LO}_{g q \to q H}}{d\xi_p}\(N,\xi_p\)=\frac{\as\Cf}{\pi}\frac{1}{\xi_p} \frac{\Gamma\(\frac{1}{2}\)\Gamma\(N\)}{\Gamma\(N+\frac{1}{2}\)}\Bigg( \,_2F_1\(\frac{1}{2},N,N+\frac{1}{2},\(\sqrt{1+\xi_p}-\sqrt{\xi_p}\)^4\)\notag\\
&-\frac{\(4+3\xi_p\)}{\(\sqrt{1+\xi_p}+\sqrt{\xi_p}\)^2}\frac{N}{N+\frac{1}{2}}\,_2F_1\(\frac{1}{2},N+1,N+\frac{3}{2},\(\sqrt{1+\xi_p}-\sqrt{\xi_p}\)^4\)\notag\\
&+3\frac{1+\xi_p}{\(\sqrt{1+\xi_p}+\sqrt{\xi_p}\)^4}\frac{N\(N+1\)}{\(N+\frac{1}{2}\)\(N+\frac{3}{2}\)} \,_2F_1\(\frac{1}{2},N+2,N+\frac{5}{2},\(\sqrt{1+\xi_p}-\sqrt{\xi_p}\)^4\)\notag\\
&-\frac{1}{\(\sqrt{1+\xi_p}+\sqrt{\xi_p}\)^6} \frac{N\(N+1\)\(N+2\)}{\(N+\frac{1}{2}\)\(N+\frac{3}{2}\)\(N+\frac{5}{2}\)} \,_2F_1\(\frac{1}{2},N+3,N+\frac{7}{2},\(\sqrt{1+\xi_p}-\sqrt{\xi_p}\)^4\)\Bigg)\\
&\,\notag\\
&\frac{d\hat{\sigma}^{\rm LO}_{q q \to g H}}{d\xi_p}\(N,\xi_p\)=\frac{2\as\Cf^2}{\pi}\frac{1}{\(\sqrt{1+\xi_p}+\sqrt{\xi_p}\)^2}\Bigg( \,_2F_1\(\frac{1}{2},N,N+\frac{1}{2},\(\sqrt{1+\xi_p}-\sqrt{\xi_p}\)^4\)\notag\\
&-2\frac{\(1+\xi_p\)}{\(\sqrt{1+\xi_p}+\sqrt{\xi_p}\)^2}\frac{N}{N+\frac{1}{2}}\,_2F_1\(\frac{1}{2},N+1,N+\frac{3}{2},\(\sqrt{1+\xi_p}-\sqrt{\xi_p}\)^4\)\notag\\
&+\frac{1}{\(\sqrt{1+\xi_p}+\sqrt{\xi_p}\)^4}\frac{N\(N+1\)}{\(N+\frac{1}{2}\)\(N+\frac{3}{2}\)} \,_2F_1\(\frac{1}{2},N+2,N+\frac{5}{2},\(\sqrt{1+\xi_p}-\sqrt{\xi_p}\)^4\)
\end{align}
\normalsize
where $\,_2F_1$ is the Hypergeometric Function.

Finally, the matching constant 
\beq
g_0\,_{ij}\(\xi_p\)=1\,+\,g^{(1)}_0\,_{ij}\(\xi_p\) \(\frac{\as}{\pi}\)+\Ord\(\as^2\)
\eeq
is given in Ref.~\cite{deFlorian:2005fzc}. We have recomputed it independently,
obtaining
\begin{align}
&g_0^{(1)}\,_{g g}\(\xi_p\)=\frac{67}{36}\Ca-\frac{5}{18}\Nf+\Ca\zeta_2-\beta_0 \ln\frac{\xi_p}{1+\xi_p}-\frac{1}{8}\Ca\ln^2\frac{\xi_p}{1+\xi_p}\notag\\
&+2\Ca \Li_2\(1-\frac{\sqrt{\xi_p}}{\sqrt{1+\xi_p}}\)+\Ca\ln\(1-\frac{\sqrt{\xi_p}}{\sqrt{1+\xi_p}}\)\ln\frac{\xi_p}{1+\xi_p}\notag\\
&-\frac{1}{2}\Ca\ln\(1+\frac{\sqrt{\xi_p}}{\sqrt{1+\xi_p}}\)\ln\frac{\xi_p}{1+\xi_p}+\frac{1}{2}\Ca\ln^2\(1+\frac{\sqrt{\xi_p}}{\sqrt{1+\xi_p}}\)+2\beta_0\ln^2\(1+\frac{\sqrt{\xi_p}}{\sqrt{1+\xi_p}}\)\notag\\
&+\Ca \Li_2\(\frac{2\sqrt{\xi_p}}{\sqrt{1+\xi_p}+\sqrt{\xi_p}}\)-\frac{\(\Ca-\Nf\)\(\sqrt{\xi_p}\sqrt{1+\xi_p}\(1+\xi_p\)-2\xi_p-\xi_p^2\)}{6\(1+8\xi_p+9\xi_p^2\)}\\
&\,\notag\\
&g_0^{(1)}\,_{g q}\(\xi_p\)=-\frac{7}{4}\Cf+\frac{134}{36}\Ca-\frac{20}{36}\Nf-8\Cf\zeta_2+12\Ca\zeta_2-4\beta_0\ln\frac{\xi_p}{1+\xi_p}+\frac{3}{2}\Cf\ln\frac{\xi_p}{1+\xi_p}\notag\\
&-\frac{1}{2}\Ca\ln^2\frac{\xi_p}{1+\xi_p}+4\(\Cf+\Ca\)\Li_2\(2,1-\frac{\sqrt{\xi_p}}{\sqrt{1+\xi_p}}\)\notag\\
&+\frac{2\(\Ca-\Cf\)\(1+3\xi_p+3\sqrt{\xi_p}\sqrt{1+\xi_p}\)}{2\sqrt{\xi_p}\sqrt{1+\xi_p}+1+3\xi_p}+8\beta_0\ln\(1+\frac{\sqrt{\xi_p}}{\sqrt{1+\xi_p}}\)\notag\\
&-3\Cf\ln\(1+\frac{\sqrt{\xi_p}}{\sqrt{1+\xi_p}}\)+2\Cf\ln\(1-\frac{\sqrt{\xi_p}}{\sqrt{1+\xi_p}}\)\ln\frac{\xi_p}{1+\xi_p}\notag\\
&+2\Ca\ln\(1-\frac{\sqrt{\xi_p}}{\sqrt{1+\xi_p}}\)\ln\frac{\xi_p}{1+\xi_p}-2\Cf\ln\(1+\frac{\sqrt{\xi_p}}{\sqrt{1+\xi_p}}\)\ln\frac{\xi_p}{1+\xi_p}\notag\\
&-2\Cf\ln^2\(1+\frac{\sqrt{\xi_p}}{\sqrt{1+\xi_p}}\)+4\Cf\Li_2\(\frac{2\sqrt{\xi_p}}{\sqrt{1+\xi_p}+\sqrt{\xi_p}}\)\\
&\,\notag\\
&g_0^{(1)}\,_{q q}\(\xi_p\)=-\frac{9}{2}\Cf+\frac{79}{12}\Ca-\frac{5}{6}\Nf+12\Cf\zeta_2-10\Ca\zeta_2-\frac{\(\Cf-\Ca\)\sqrt{1+\xi_p}}{\sqrt{\xi_p}}\notag\\
&+4\Cf\Li_2\(1-\frac{\sqrt{\xi_p}}{\sqrt{1+\xi_p}}\)-\frac{3}{4}\Cf\ln\frac{\xi_p}{1+\xi_p}-\beta_0\ln\frac{\xi_p}{1+\xi_p}+\frac{1}{4}\Ca\ln^2\frac{\xi_p}{1+\xi_p}\notag\\
&-\frac{1}{2}\Cf\ln^2\frac{\xi_p}{1+\xi_p}+2\Cf\ln\(1-\frac{\sqrt{\xi_p}}{\sqrt{1+\xi_p}}\)\ln\frac{\xi_p}{1+\xi_p}+\frac{3}{2}\Cf\ln\(1+\frac{\sqrt{\xi_p}}{\sqrt{1+\xi_p}}\)\notag\\
&+2\beta_0\ln\(1+\frac{\sqrt{\xi_p}}{\sqrt{1+\xi_p}}\)+\Ca\ln^2\(1+\frac{\sqrt{\xi_p}}{\sqrt{1+\xi_p}}\)\notag\\
&-\Ca\ln\(1+\frac{\sqrt{\xi_p}}{\sqrt{1+\xi_p}}\)\ln\frac{\xi_p}{1+\xi_p}+2\Ca\Li_2\(\frac{2\sqrt{\xi_p}}{\sqrt{1+\xi_p}+\sqrt{\xi_p}}\).
\end{align}

\subsection{Transverse momentum resummation}
\label{subapp:smallpt}

We collect expressions of the coefficients
which determine our generalized transverse momentum resummation
$\frac{d\hat{\sigma}_{ij}^{\rm tr'}}{d\xi_p}$
Eq.~(\ref{eq:smallptjoint}) for  Higgs  production
in gluon fusion up to NNLL order. 
They are based on the results of
Refs.~\cite{Catani:2000vq,Catani:2011kr,Bozzi:2008bb,Catani:2013tia,Catani:2015vma,Bozzi:2005wk}) 
for the standard transverse momentum resummation
$\frac{d\hat{\sigma}_{ij}^{\rm tr}}{d\xi_p}$
Eq.~(\ref{eq:smallptjoint3}).  As shown in those references, the functions 
$A^{\pt}$ and $B^{\pt}$ in Eq.~(\ref{eq:sudpt}) depend on
the leading-order partonic subprocess; they will thus be labeled by
a subscript $g$ to denote the gluon channel.

The function $A^{\pt}_g$ in Eq.~\eqref{eq:smallptjoint} is a series of
constants:
\beq
A^{\pt}_g\(\as\)=A_g^{\pt, (1)} \(\frac{\as}{\pi}\)+A_g^{\pt, (2)} 
\(\frac{\as}{\pi}\)^2+A_g^{\pt, (3)} \(\frac{\as}{\pi}\)^3+ \Ord\(\as^4\),
\eeq
where
\begin{align}
&A_g^{\pt, (1)}=\Ca\\
&A_g^{\pt, (2)}=\frac{\Ca}{2}\left[\Ca\(\frac{67}{18}-\zeta_2\)-\frac{5}{9}\Nf\right]\\
&A_g^{\rm \pt, (3)}=\frac{\Ca}{4}\Bigg[\Ca^2\(\frac{15503}{648}-\frac{67}{9}\zeta_2-11\zeta_3+\frac{11}{2}\zeta_4\)+\Cf\Nf\(-\frac{55}{24}+2\zeta_3\)\notag\\
&\qquad\qquad+\Ca\Nf\(-\frac{2051}{324}+\frac{10}{9}\zeta_2\)-\frac{25}{81}\Nf^2\Bigg].
\end{align}

The function  $\mathcal{B}_g\(N\)$, is a two by two matrix, since PDFs
evolution only involves the singlet sector in the case of Higgs  production.  It includes all exponentiated
terms which
vanish as $N\to\infty$ and  admits the following expansion in $\as$
\begin{align}
&\mathcal{B}_g\(N,\as\)=\mathcal{B}_g^{(1)}\(N\) \(\frac{\as}{\pi}\)+\mathcal{B}_g^{(2)}\(N\)  \(\frac{\as}{\pi}\)^2 +\Ord\(\as^3\),\notag\\
&\,\notag\\\label{eq:calB1}
&\mathcal{B}_g^{(1)}\(N\)=2\gamma^{\rm reg, (1)} \(N\)\\\label{eq:calB2}
&\mathcal{B}_g^{(2)}\(N\)=\tilde{B}^{\rm \pt, (2)}+2\gamma^{\rm reg, (2)}\(N\)
\end{align}
where both $\gamma^{\rm reg, (i)}$ and $\tilde{B}^{\rm \pt, (2)}$ are two by two matrices defined as
\begin{align}
\label{eq:gammareg1}
&\gamma_{jk}^{\rm reg, (i)}\(N\)=\gamma_{jk}^{(i)}\(N\) \qquad \mbox{ if } j\neq k \neq g\\
\label{eq:gammareg2}
&\gamma_{gg}^{\rm reg, (i)}\(N\)=\gamma_{gg}^{(i)}\(N\)+A_g^{\rm th, (i)} \(\ln N +\gammae\)-\delta P_{gg}^{i}\\
&\notag\\
&\tilde{B}_{jk}^{\rm \pt, (2)}=0 \qquad \mbox{ if } j\neq k \neq g\\
&\tilde{B}_{gg}^{\rm \pt, (2)}=\beta_0 \Ca \zeta_2
\end{align}
with $\Ca=3, \Cf=\frac{4}{3}$. Here, $\gamma^{(i)}$ are the coefficients
of $\(\frac{\as}{\pi}\)^i$ in the expansions of
the Altarelli-Parisi anomalous dimensions,  the cusp anomalous dimension $A^{\rm th,
  (i)}$  was given in Eqs.~(\ref{eq:cusp}-\ref{eq:cusp3}) and
$\delta P_{gg}^{(i)}$ is the coefficient
of $\(\frac{\as}{\pi}\)^i\delta\(1-z\)$ in 
the expansion of the splitting function $P_{gg}(z)$.
From Eqs.~(\ref{eq:gammareg1}-\ref{eq:gammareg2}), it is easy to see that
\beq
\gamma^{\rm reg, (i)}\(N\)=\Ord\(\frac{1}{N}\)
\eeq
at large $N$.

The function
$\mathcal{H}_{ij}\(N,\as\(M^2\)\)$ is given in
Refs.~\cite{Bozzi:2005wk,Catani:2011kr}, and it is defined factoring out 
the inclusive cross section  Eq.~(\ref{eq:sigma0}).
The coefficients which control logarithmically enhanced contributions
as $N\to\infty$ are 
\begin{align}
&D_g^{\pt}\(\as\)=D_g^{\pt, (1)}\(\frac{\as}{\pi}\)+D_g^{\pt, (2)}\(\frac{\as}{\pi}\)^2+\Ord\(\as^3\)\notag\\
&\,\notag\\
&D_g^{\pt,(1)}=0,\\
&D_g^{\pt, (2)}=\Ca^2\(-\frac{101}{27}+\frac{7}{2}\zeta_3\)+\frac{14}{27}\Ca\Nf
\end{align}
while the term $\mathcal{H}_{ij}^0\(\as\)$ (constant as $N\to\infty$) is
\begin{align}
&\mathcal{H}_{ij}^0\(\as\)=0 \qquad \mbox{ if $i\neq j\neq g$}\notag\\
&\mathcal{H}_{gg}^0\(\as\)=1+\mathcal{H}_{gg}^{0, (1)} \(\frac{\as}{\pi}\)+ \mathcal{H}_{gg}^{0, (2)} \(\frac{\as}{\pi}\)^2 + \Ord\(\as^3\)\notag\\
&\,\notag\\
&\mathcal{H}_{gg}^{0, (1)}=3\Ca \zeta_2,\\
&\mathcal{H}_{gg}^{0, (2)}=\Ca^2\(\frac{93}{16}+\frac{67}{12}\zeta_2-\frac{55}{18}\zeta_3+\frac{65}{8}\zeta_4\)+\Ca\Nf\(-\frac{5}{3}-\frac{5}{6}\zeta_2-\frac{4}{9}\zeta_3\).
\end{align}
Finally, terms which vanish as $N\to\infty$
are~\cite{Bozzi:2005wk,Catani:2011kr} 
\begin{align}
\mathcal{H}_{gg}^{\rm f.o., (1)}\(N\)&=0\\
\mathcal{H}_{gq}^{\rm f.o., (1)}\(N\)&=\mathcal{H}_{qg}^{\rm (1)}\(N\)=\frac{1}{2}\Cf \frac{1}{N+1}.
\end{align}
The  complicated $\Ord(\alpha_s^2)$  coefficients  can be obtained by
Mellin transform of coefficients given in
Ref.~\cite{Catani:2011kr} :
\begin{align}
\mathcal{H}_{ij}^{\rm f.o, (2)}\(N\)&= \int_0^1 dz\,z^{N-1}
\mathcal{H}^{H(2)}_{gg \leftarrow ij}\(z\),\quad ij=qq,qg\\
\mathcal{H}_{gg}^{\rm f.o, (2)}\(N\)&=-\mathcal{H}_{gg}^{0, (2)}+D_g^{\pt, (2)} \ln N + \int_0^1 dz\,z^{N-1} \mathcal{H}^{H(2)}_{gg \leftarrow gg}\(z\),
\end{align}
with $\mathcal{H}^{H(2)}_{gg \leftarrow ij}$ given in Eqs.~(22-24) of Ref.~\cite{Catani:2011kr}.

\subsection{Threshold resummation for inclusive cross section}
\label{subapp:inclN}
We provide here the remaining coefficients which fully determine
NNLL* threshold resummation at the inclusive
level~\cite{Moch:2005ba,Catani:2001ic}. This can written as
\beq\label{eq:thrres1}
\hat{\sigma}^{\rm res}\(N\)=\sigma_0\,H_{gg}^{\rm th}\,\exp\left[G\(N\)\right]
\eeq
with
\beq\label{eq:thrres2}
G\(N\)=\int_0^1 dz\,\frac{z^{N-1}-1}{1-z}\left[2\int_{M^2}^{\(1-z\)^2 M^2} \frac{dq^2}{q^2} A^{\rm th}_g\(\as\(q^2\)\)+D_g^{\rm th}\(\as\(\(1-z\)^2 M^2\)\)\right].
\eeq

The the cusp anomalous dimension $A^{\rm th,
  (i)}$  was given in Eqs.~(\ref{eq:cusp}-\ref{eq:cusp3}). 
The  function $D_g\(\as\(\(1-z\)^2 M^2\)\)$, which
contains  contributions from large-angle (non collinear) 
soft gluon emission, is given by
\begin{align}
&D_g^{\rm th}\(\as\)=D^{\rm th, (1)}_g\(\frac{\as}{\pi}\)+D^{\rm th, (2)}_g\(\frac{\as}{\pi}\)^2 + \Ord\(\as^3\)\notag\\
&\quad\notag\\\label{eq:largang1}
&D_g^{\rm th, (1)}=0,\\\label{eq:largang2}
&D_g^{\rm th, (2)}=\Ca\(\Ca\(-\frac{101}{27}+\frac{11}{3}\zeta_2+\frac{7}{2}\zeta_3\)+\Nf\(\frac{14}{27}-\frac{2}{3}\zeta_2\)\)
\end{align}
and the hard function is
\label{eq:appHth}
\begin{align}
&H_{gg}^{\rm th}\(as\)=1+H_{gg}^{\rm th, (1)}\(\frac{\as}{\pi}\)+\Ord\(\as^2\)\notag\\
&\quad\notag\\
&H_{gg}^{\rm th, (1)}= 4\Ca\zeta_2.
\end{align}

\phantomsection

\end{document}